\documentclass[showpacs,prl,onecolumn,aps,superscriptaddress,preprintnumbers,nofootinbib]{revtex4}
\usepackage[T1]{fontenc}
\usepackage[latin9]{inputenc}
\usepackage{amsmath,amssymb,bigints}
\usepackage{epsfig}
\usepackage{graphicx}
\usepackage{amsmath}
\usepackage{amsfonts}
\def\slashchar#1{\setbox0=\hbox{$#1$}     		
   \dimen0=\wd0                                 	
   \setbox1=\hbox{/} \dimen1=\wd1               	
   \ifdim\dimen0>\dimen1                        	
      \rlap{\hbox to \dimen0{\hfil/\hfil}}      	
      #1                                        	
   \else                                        	
      \rlap{\hbox to \dimen1{\hfil$#1$\hfil}}   	
      /                                         	
   \fi}

\renewcommand{\vec}{\boldsymbol}
\newcommand{\beq}{\begin{equation}}
\newcommand{\eeq}{\end{equation}}
\newcommand{\bea}{\begin{eqnarray}}
\newcommand{\eea}{\end{eqnarray}}
\newcommand{\baa}{\begin{array}}
 \newcommand{\eaa}{\end{array}}
\renewcommand{\vec}[1]{\boldsymbol{#1}}
\def\eq#1{{Eq.~(\ref{#1})}}
\def\fig#1{{Fig.~\ref{#1}}}

\newcommand{\bas}{\bar{\alpha}_S}
\newcommand{\as}{\alpha_S}

\newcommand{\nn}{\nonumber}

\newcommand{\h}{\frac{1}{2}}

\newcommand{\ga}{\gamma}

\newcommand{\Lb}{\left(}
\newcommand{\Rb}{\right)}

\renewcommand{\vec}[1]{\boldsymbol{#1}}

\begin{document}

\title{ Gribov-Zwanziger confinement, high energy evolution
 and large impact
 parameter behaviour of the scattering amplitude}
\author{E. ~Gotsman}
\email{gotsman@post.tau.ac.il}
\affiliation{Department of Particle Physics, School of Physics and Astronomy,
Raymond and Beverly Sackler
 Faculty of Exact Science, Tel Aviv University, Tel Aviv, 69978, Israel}
 \author{E.~ Levin}
\email{leving@tauex.tau.ac.il, eugeny.levin@usm.cl}
\affiliation{Department of Particle Physics, School of Physics and Astronomy,
Raymond and Beverly Sackler
 Faculty of Exact Science, Tel Aviv University, Tel Aviv,
 69978, Israel}\affiliation{ Departamento de F\'\i sica,
Universidad T$\acute{e}$cnica Federico Santa Mar\'\i a   and
Centro Cient\'\i fico-Tecnol$\acute{o}$gico de Valpara\'\i so,
Casilla 110-V,  Valparaiso, Chile}

\date{\today}


\begin{abstract}

In this paper we derive the high energy evolution equation  in 
 the  Gribov-Zwanziger 
approach, for the  confinement of quarks and gluons. We demonstrate 
that the
 new equation generates  an  exponential decrease of the scattering
 amplitude at large impact parameter, and resolves the main difficulties
 of CGC (Colour Glass Condensate)  high energy effective theory. 
  Such behaviour   occurs,
  if the gluon propagator in  Gribov-Zwanziger 
approach, does not vanish at small momenta.  Solving the non-linear 
equation
 for deep inelastic scattering, we show that the suggested equation leads 
to a
 Froissart disc with  radius ($R_F$),   which increases as
 $ R_F \,\propto Y = \ln (1/x)$,  and with a finite width for the
 distribution over $| b - R_F|$.
 \end{abstract}
\preprint{TAUP - 3040/19}
\maketitle


\tableofcontents
\newpage

\section{ Introduction}
It is well known that the Balitsky-Kovchegov equation\cite{BK}: 
\bea 
\frac{\partial}{\partial Y} N\Lb \vec{r}, \vec{b} ,  Y \Rb &=&\bas\!\! \int \frac{d^2 \vec{r}'}{2\,\pi}\,K\Lb \vec{r}', \vec{r} - \vec{r}'; \vec{r}\Rb \Bigg\{N\Lb \vec{r}',\vec{b} - \h \Lb \vec{r} - \vec{r}' \Rb, Y\Rb + 
N\Lb\vec{r} - \vec{r}', \vec{b} - \h \vec{r}', Y\Rb \,\,- \,\,N\Lb \vec{r},\vec{b},Y \Rb\nn\\
& & ~~~~~~~- N\Lb\vec{r} - \vec{r}', \vec{b} - \h \vec{r}', Y\Rb   \, N\Lb \vec{r}',\vec{b} - \h \Lb \vec{r} - \vec{r}' \Rb, Y\Rb\Bigg\} \label{BK}
\eea
generates a scattering amplitude which 
 decreases  as a power of $b$ at large impact parameter  
 (see Ref.\cite{KOLEB} for review). 
 
In \eq{BK}  the kernel $K\Lb \vec{r}', \vec{r} - \vec{r}'; \vec{r}\Rb$
 describes
 the decay of the dipole of size $r$,
into two dipoles with  sizes $r'$ and $| \vec{r}\,-\,\vec{r}'|$,
 respectively. It has the form:

\beq \label{KERC}
K\Lb \vec{r}', \vec{r} - \vec{r}'; \vec{r}\Rb \,\,=\,\,\frac{r^2}{r'^2\,\Lb \vec{r} - \vec{r}'\Rb^2}
\eeq

  Indeed, at large $b$ we can neglect the non-linear
 term in \eq{BK}, and  the linear BFKL equation\cite{FKL,BFKL} 
determines
 the large $b$ behaviour. It is known that the eigenfunction of this equation
 (the scattering amplitude of two dipoles with sizes $r$ and $R$) has the
 following form \cite{LIP}
\beq \label{EIGENF}
\phi_\gamma\Lb \vec{r} , \vec{R}, \vec{b}\Rb\,\,\,=\,\,\,\Lb \frac{
 r^2\,R^2}{\Lb \vec{b}  + \h(\vec{r} - \vec{R})\Rb^2\,\Lb \vec{b} 
 -  \h(\vec{r} - \vec{R})\Rb^2}\Rb^\gamma\,\,\xrightarrow{b\,\gg\,
r,R}\,\,\Lb \frac{ r^2\,R^2}{b^4}\Rb^\gamma\,\,\equiv\,\,e^{\gamma\,\xi}
~~\mbox{with}~~ \xi\,=\,\ln \Lb \frac{ r^2\,R^2}{b^4}\Rb
\eeq
 
 \eq{EIGENF} shows the power-like decrease at large $b$, which 
 leads  to  the violation of the Froissart theorem\cite{FROI} 
generating  a
 cross section, which at high energies increases as a power of energy   
\cite{KW,FIIM}. The solution of this problem requires
  introducing a new  dimensional scale.  A variety of ideas to 
overcome this problem
 have been suggested in Refs. \cite{LERYB1,LERYB2,LETAN,QCD2,KHLEP,KKL,
FIIM,GBS1,BLT,GKLMN,HAMU,MUMU,BEST1,BEST2,KOLE,LETA,LLS,LEPION,KHLE,KAN,
GOLEB}.  In this paper we intend to use the Gribov-Zwanziger 
 approach\cite{GRI0,GRI1,GRI2,GRI3,GRI4,GRREV,DOKH}  for  the
 confinement of quarks and gluons. In particular, we will use
 the Gribov gluon propagator in a form which describes the recent 
lattice QCD estimates \cite{DOS}.
 
 The plan of this paper is as follows. In the next section we illustrate 
 the problem of  the large impact parameter behaviour of the BK equation 
 as an 
 example of the first iterations of this equation. In section III we 
discuss
 the model: non-abelian gauge theories with Higgs mechanism of mass
 generation, which has been suggested in Ref.\cite{LLS} for  high
 energy scattering. In spite of the fact that this model does not have
 the confinement of quarks and gluons, we found  it instructive to 
reproduce
 the large impact parameter behaviour of the scattering amplitude in
 this model. It should be stressed that this model not only leads to
 the exponential decrease of the scattering amplitude at large $b$, but
 has the same spectrum of energies as the massless BFKL equation in
 QCD. Section IV is the  key  chapter in the paper. It  contains 
a discussion of 
  the modification of the BFKL evolution equation for the
 Gribov-Zwanziger  approach, to the confinement problem. We show that
  this mechanism of confinement   introduces  a   new dimensional 
parameter,
 and  it leads to the exponential decrease of the scattering amplitude at
 large $b$, only if the gluon propagator does not vanish at zero momentum.
 In other words, we need to introduce two dimensional parameters to provide
 the correct large $b$ behaviour in the framework of  the Gribov-Zwanziger 
 approach to  confinement. In section V we
 discuss the non-linear equation with a generalized kernel, and show 
that
 this equation generates the Froissart- type behaviour of the scattering
 amplitude with a radius which increase as $Y = \ln(1/x)$.
  Finally, in Section VI we discuss our results and future prospects.  


\section{Iterations of BK equation}

We start illustrating the problem of large $b$ behaviour with the first 
iteration of \eq{BK}. At large $b$, we can neglect the non-linear term and
 concentrate our efforts on solution of linear  BFKL \cite{FKL,BFKL}
 equation .
The general initial condition   $N^{(0)}$  generates the Green's function 
in
 the impact  parameter representation, and   has the following form:
\beq \label{IC}
N^{(0)}\,\,=\,\, r^2\,\delta^{(2)}\Lb \vec{b}\Rb
\eeq
\eq{IC} gives the dimensionless  function which plays the role of the
 Green's function  in $b$-space.

 Plugging  this initial condition in \eq{BK}, one can see that we 
obtain
 the first iteration in the form:
\beq \label{FEQ3}
N^{(1)}\Lb \vec{r}, \vec{b} ,  Y \Rb\,\,=\,\,\bas\,Y\,\Bigg\{ \frac{1}{\pi}\,\frac{r^2}{\Lb 2\,\vec{b}\Rb^2}\,\,
 \,-\,\ln r^2\,\,N^{(0)}\Lb \vec{r},\vec{b}\Rb\Bigg\}
\eeq	

Therefore, one can see that the initial conditions which have a sharp
 decrease in $b$, generate the power-like dependence of the solution
 to the BFKL equation. The next iteration leads to
\beq \label{FEQ4}
N^{(2)}\Lb \vec{r}, \vec{b} ,  Y \Rb\,\,=\,\,\h \Lb \bas \,Y\Rb^2\int d^2 r' 2\,\frac{r^2}{\Lb \vec{r}\,-\,\vec{r}'\Rb^2}
\frac{1}{\Lb 2\,\vec{b}\Rb^2}\,\,\xrightarrow{b \gg r' \gg r}\,\,\h \Lb \bas \,Y\Rb^2\,\,\int^{b^2}_{r^2}
 \frac{d r'^2}{r'^2} \frac{r^2}{b^2}
 \eeq

Therefore, the BFKL equation generates the power-like decrease  of the
 scattering amplitude in the first iteration, while in the following 
iterations
 the typical $r'$ turns out to be much smaller than $b$. 
 
 It is instructive to recall that the power-like decrease, which comes
 from the integration over $r'$ in \eq{FEQ3}, corresponds to the gluon
 reggeization term in the  momentum representation. Indeed, in the 
momentum
 representation the BFKL equation takes the form\cite{FKL,BFKL}:
 
\bea \label{BFKLM}
\frac{\partial N\Lb Y; q, Q_T\Rb}{ \partial Y}\,\,&=&\,\,\bas
 \Bigg(\int \frac{d^2 q'}{2 \pi}  K_{\rm em}\Lb \vec{q} - \vec{q}',
 \vec{Q}_T\Rb N\Lb Y; q', Q_T\Rb\,\,-\,\,K_{\rm reg}\Lb \vec{q}
 - \vec{q}', \vec{Q}_T\Rb N\Lb Y; q, Q_T\Rb\Bigg)\nn\\
 &=& \bas \,\int \frac{d^2 q'}{2 \pi}  K_{\rm em}\Lb \vec{q} - \vec{q}',
 \vec{Q}_T\Rb N\Lb Y; q', Q_T\Rb\,\,-\,\,\omega_G\Lb q'\Rb  \,\,-\,\,\omega_G\Lb \vec{Q}_T\,-\,\vec{q'}\Rb\nn\\ \,\,&=& \,\,\bas
\int \frac{d^2 q'}{2 \pi}  K\Lb \vec{q} - \vec{q}',
 \vec{Q}_T\Rb N\Lb Y; q', Q_T\Rb 
  \eea

where $\bas = \Lb N_c/\pi\Rb \as$.  The kernel $K_{\rm em}$ describes
 the emission of a gluon, while  kernel $K_{\rm reg}$ is responsible
 for the reggeization of gluons in t-channel.
They have the forms:
\bea \label{KERM}
K_{\rm em}\Lb \vec{q} - \vec{q}', \vec{Q}_T\Rb\,\,&=&\,\,\h \frac{1}{\Lb \vec{q} - \vec{q}'\Rb^2} \Bigg\{ - \frac{Q^2_T 
\Lb \vec{q} - \vec{q}'\Rb^2 }{\Lb \vec{Q}_T - \vec{q}'\Rb^2\,q'^2}  \,+1\,+ \frac{\Lb \vec{Q}_T - \vec{q}\Rb^2\,q'^2}{\Lb \vec{Q}_T - \vec{q}'\Rb^2\,q^2}\Bigg\} \xrightarrow{Q_T =0} \frac{1}{\Lb \vec{q} - \vec{q}'\Rb^2} \\
K_{\rm reg}\Lb \vec{q} - \vec{q}', \vec{Q}_T\Rb\,\,&=&
\h \frac{1}{\Lb \vec{q} - \vec{q}'\Rb^2} \Bigg\{  \frac{q^2 
}{\Lb \vec{q} - \vec{q}'\Rb^2\,+\,q'^2}\,\,+\,\, \frac{\Lb \vec{Q}_T \,-\,\vec{q}\Rb^2 
}{\Lb \vec{q} - \vec{q}'\Rb^2\,+\,\Lb \vec{Q}_T - \vec{q}'\Rb^2}\Bigg\} \xrightarrow{Q_T =0}\frac{1}{\Lb \vec{q} - \vec{q}'\Rb^2}   \frac{q^2 
}{\Lb \vec{q} - \vec{q}'\Rb^2\,+\,q'^2}\nn
\eea 
 $K\Lb \vec{q} - \vec{q}',\vec{Q}_T\Rb $ is equal to
 
 \beq \label{KK}
 K\Lb \vec{q} - \vec{q}',
 \vec{Q}_T\Rb N\Lb Y; q', Q_T\Rb\,\,=\,\,  K_{\rm em}\Lb \vec{q} - \vec{q}',
 \vec{Q}_T\Rb N\Lb Y; q', Q_T\Rb \,\,-\,\,K_{\rm reg}\Lb \vec{q} - \vec{q}',
 \vec{Q}_T\Rb N\Lb Y; q, Q_T\Rb 
 \eeq 
 The gluon trajectory $\omega_G\Lb q\Rb$ is equal to
 
 \beq \label{OMEGL}
 \omega_G\Lb q \Rb \,\,=\,\,\bas\,\int \frac{d^2\,q'}{4\,\pi} \,\frac{q^2}{q'^2\,\Lb \vec{q}\,-\,\vec{q}'\Rb^2}
 \eeq

 The reggeization term of \eq{BFKLM} leads  to the 
 power-like behaviour at large impact parameter.
Therefore, we need to understand, what type of non-perturbative
 corrections could change this reggeization kernel to 
provide the exponential decrease, of the scattering amplitude at
 large impact parameters.

 
  \section{The model: nonabelian gauge theories with the Higgs
 mechanism for  mass generation.}
 
 
  \subsection{BFKL equation}
 
\begin{figure}[ht]
     \begin{center}
     \includegraphics[width=14cm]{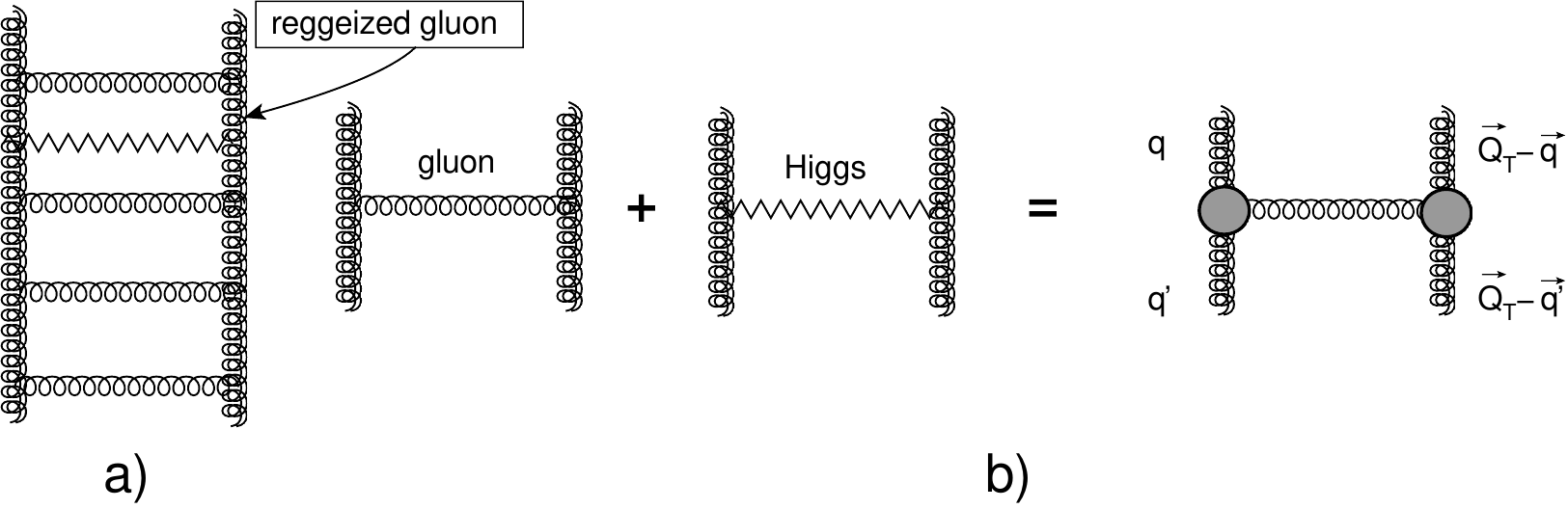} 
     \end{center}    
      \caption{ The massive BFKL equation (\protect\fig{eq}-a) and its
 kernel (\protect\fig{eq}-b)  }
\label{eq}
   \end{figure}

 
In this section we wish to answer the question: what  is  the large
 impact parameter behaviour in the non-abelian 
Yang-Mills theories with a Higgs particle. In these theories, we  
introduce 
the
 non-perturbative scale as the mass of Higgs, and we would like to see
 how this dimensional scale manifests itself in the large $b$ behaviour
 of the scattering amplitude. 
It was shown by Fadin, Lipatov and Kuraev \cite{FKL},  that  the high
 energy amplitude satisfies the BFKL equation (see \fig{eq}) which has
 been written for  colour $SU(N_c) $, ($N_c$ is the number of colours)
 with the Higgs mechanism of mass generation, in Ref.\cite{LLS}.

It has the form of \eq{BFKLM},  with the kernels that have the following 
forms:
\begin{subequations}
\bea 
K_{\rm em}\Lb \vec{q} - \vec{q}', \vec{Q}_T\Rb\,\,&=&  \frac{1}{\Lb \vec{q} \,-\,\vec{q}'\Rb^2 + m^2}\Bigg(\frac{q^2_1 + m^2}{q'^2 + m^2}\,+\,\frac{\Lb \vec{Q}_T \,-\,\vec{q}\Rb^2 + m^2}{ \Lb \vec{Q}_T \,-\,\vec{q}'\Rb^2+ m^2}\Bigg)\,\,-\,\,\frac{q^2 +\frac{N^2_c+1}{N^2_c} m^2}{(q'^2 + m^2) ( \Lb \vec{Q}_T \,-\,\vec{q}\Rb^2 + m^2)}\;\label{KEREMM}\\
\omega_G\Lb q \Rb \,\,&=&\,\, \int \frac{ d^2 q' }{4\,\pi}\frac{\Lb q^2 + m^2\Rb}{\Lb q'^2 + m^2\Rb \Lb\Lb  \vec{q} - \vec{q}'\Rb^2 + m^2 \Rb}=\,\,  \h \frac{q^2 +m^2}{|q|\sqrt{q^2 + 4  m^2}}\ln \frac{\sqrt{q^2 + 4 m^2} + |q|}{\sqrt{q^2 + 4 m^2} - |q|}  \label{GLTR}
  \eea   
  \end{subequations}

  As one can see from \eq{GLTR}, $\omega_G\Lb q\Rb$ has singularities
 at $q^2\,=\, - 4 \,m^2$, which will generate the exponential
 decrease of the scattering amplitude at large $b$. As we have
 mentioned, the reggeization terms in coordinate representation
 generate the first term in \eq{BK}. Using formulae {\bf 8.411(1),
  8.411(7)} and {\bf 6.532(4)} of Ref. \cite{RY}) we obtain
\beq \label{CORE1}
\int \frac{d^2 q'}{2 \pi} \frac{e^{i \vec{r} \cdot \vec{q}'}}{ q'^2 + m^2}\,=\,\int^{+ \infty}_{-\infty}\frac{q' d q'\,J_0\Lb r q'\Rb}{q'^2 + m^2 }\,\,=\,\,K_0\Lb r m\Rb
\eeq
where $J_0\Lb z\Rb$ and $K_0\Lb z\Rb$ are the Bessel functions of the
 first and second kinds, respectively (see Ref. \cite{RY}).  
  Bearing \eq{CORE1} in mind, one can see that coordinate image  of the
 gluon trajectory can be written as follows:
  \begin{subequations}  
  \bea
   \omega_G\Lb r\Rb\,\,\,&=&\,\,\h\,\Lb - \nabla^2\,\,+\,\,m^2\Rb \,K^2_0\Lb r\, m\Rb \,\,=\,\, - \,\,\h\,\,m^2\Big( 2 K^2_1\Lb m\,r\Rb\,+\,
  K^2_0\Lb m\,r\Rb\Big)  \label{OMGCOR}\\
    &\xrightarrow{m\,\to\,0}&   -\frac{1}{r^2}; \label{OMGM0}\\
 &\xrightarrow{r\,\to\,0}& -\frac{1}{r^2} \,\,+\,\,\h\,m^2 \Big( 2 - \Lb C \,+\,\ln\Lb\frac{m\,r}{2}\Rb\Rb^2\Big); \label{OMGSR}\\
 &\xrightarrow{r\,\to\,\infty}& \,e^{- 2 \,m\,r}\Big(- \frac{3\,m\,\pi}{4\,r}\,\,-\,\,\frac{5\,\pi}{16\,r^2}\Big);\label{OMGLR}
   \eea
\end{subequations}    
  where $K_1\Lb z\Rb$  is the Bessel functions of the second kind and
 $C$ is the Euler constant. 
  
  The emission term of  BFKL equation in coordinate representation(the
 first two terms in \eq{BK}) have the following form:
  \beq \label{BFKLECR}
\frac{\partial}{\partial Y} N\Lb \vec{r}, \vec{b} ,  Y \Rb \,\,=\,\,-\bas\!\! \int \frac{d^2 \vec{r}'}{2\,\pi}\,\omega_G\Lb \vec{r} - \vec{r}'\Rb N\Lb \vec{r}',\vec{b} - \h \Lb \vec{r} - \vec{r}' \Rb, Y\Rb
  \eeq

  We need to add the contribution of  $K_{\rm em}\Lb \vec{q} - \vec{q}',
 \vec{Q}_T\Rb$ in the coordinate representation,   which leads to the
 term proportional to $N\Lb \vec{r}, \vec{b} ,  Y \Rb$.   Finally, the
 BFKL equation in the coordinate representation has the form:
   \beq \label{BFKLCR}
\frac{\partial}{\partial Y} N\Lb \vec{r}, \vec{b} ,
  Y \Rb \,\,=\,\,-\bas\!\! \int \frac{d^2 \vec{r}'}{2\,\pi}\,\omega_G\Lb
 \vec{r} - \vec{r}'\Rb \Bigg\{N\Lb \vec{r}',\vec{b} - \h \Lb \vec{r}
 - \vec{r}' \Rb, Y\Rb\,\,-\,\,\h N\Lb \vec{r}, \vec{b} ,  Y \Rb\Bigg\} 
 \eeq

 
  \subsection{First Iterations}
 
  Using the initial conditions of \eq{IC}, one can see that the first
 iteration of \eq{BFKLECR} leads to the following expression for large $b$:
  
  \bea \label{FI1}
N^{(1)}\Lb \vec{r}, \vec{b} , Y \Rb& =&\ -\,\bas\,Y\,\frac{1}{\pi}\,\,\int d^2 r'\,\,\omega_G\Lb \vec{r}\,-\,\vec{r}'\Rb\, N^{(0)}\Lb \vec{r}', \vec{b} \,-\,\h\Lb \vec{r} \,-\,\vec{r}'\Rb\Rb\,\,\nn\\
&=&\,\,-\,\bas\,Y\,\frac{1}{\pi}\,\,\omega_G\Lb 2 \vec{b}\Rb\,4\,b^2\,\,=\,\, 6\,\bas\,Y\, m\,b\, e^{- 4\,m\,b}
\eea

  The second iteration gives
   \bea \label{FI2}
& =&\ -\,\bas\,Y\,\frac{1}{\pi}\,\,\int d^2 r'\,\,\omega_G\Lb \vec{r}\,-\,\vec{r}'\Rb\, N^{(1)}\Lb \vec{r}', \vec{b} \,-\,\h\Lb \vec{r} \,-\,\vec{r}'\Rb\Rb\,\,\nn\\
&\xrightarrow{ \vec{b} \,\to\,\h\Lb \vec{r} \,-\,\vec{r}'\Rb}&\,\,-\,\h \Lb \bas\,Y\Rb^2\,\frac{1}{\pi}\,\,\omega_G\Lb 2 \vec{b}\Rb\,4\,b^2\,\nn\\
+ &\xrightarrow{ |\vec{r} - \vec{r}' | \approx 1/m}&\,\, \h \Lb \bas\,Y\Rb^2\,\frac{1}{\pi}\,\,\omega_G\Lb  |\vec{r} - \vec{r}' |\Rb \underbrace{ 6\,\bas\,Y\, m\,b\, e^{- 4\,m\,b}}_{N^{(1)}\Lb \vec{r}, \vec{b}\,\gg\,\vec{r}, Y \Rb}
\eea   

The  modified BFKL equation leads to the exponential decrease of
 the scattering amplitude $N\Lb \vec{r}, \vec{b} , Y \Rb$ at large $b$
 ($ m\,b\,\,\gg\,\,1$).

 
  \subsection{Solution at large impact parameter}
 
We  solve \eq{BFKLCR} at large $b$, assuming that the amplitude $N$
 has  the form:

\beq \label{SOLB1}
N\Lb \vec{r}, \vec{b} , Y \Rb\,\,\,=\,\,\,\tilde{N}\Lb r,  Y \Rb\,\,B\Lb b \Rb
\eeq
We have seen that first two iterations reproduce this form, as well as
 the eigenfunction of the BFKL equation (see 
\eq{EIGENF}). From our experience with the first iteration, we infer
 that there are two regions of integration that contribute to the
 asymptotic behaviour at large $b$:   $| \vec{b} \,-\,\h( \vec{r} \,-\,\vec{r}')| \,\to \,0$ and $
|  \vec{r} \,-\,\vec{r}'|\,\propto 1/m$.

Plugging \eq{SOLB1} into \eq{BFKLCR} we obtain the following equation:

   \beq \label{BFKLCRB}
\frac{\partial\,\tilde{N}\Lb r, Y \Rb}{\bas \,\partial Y}  B\Lb b \Rb\,\,=\,\,\underbrace{  - \,\omega_G\Lb 2\,b\Rb   \!\! \int \frac{d^2 r'}{2\,\pi} N\Lb r', b=0 , Y\Rb}_{| \vec{b} \,-\,\h( \vec{r} \,-\,\vec{r}')| \,\to \,0} \,\,
- \,\,\underbrace{\int \frac{d^2 \vec{r}'}{2\,\pi}\,\omega_G\Lb \vec{r} - \vec{r}'\Rb \Bigg\{\tilde{N}\Lb r', Y\Rb\,\,-\,\,\h\tilde{N}\Lb r , Y \Rb\Bigg\} }_{|  \vec{r} \,-\,\vec{r}'|\,\propto 1/m}\,B\Lb b \Rb
\eeq
First, we need to solve the  homogeneous     equation:
\beq \label{BFKLTI}
\frac{\partial\,\tilde{N}\Lb r, Y \Rb}{\bas \,\partial Y} \,\,=\
- \,\,\int \frac{d^2 \vec{r}'}{2\,\pi}\,\omega_G\Lb \vec{r} - \vec{r}'\Rb \Bigg\{\tilde{N}\Lb r', Y\Rb\,\,-\,\,\h\tilde{N}\Lb r , Y \Rb\Bigg\} \eeq
which in $\omega$-representation:
\beq \label{OMR}
\tilde{N}\Lb r, Y \Rb\,\,=\,\,\int^{\epsilon\,+\,i\,\infty}_{\epsilon\,-\,i\,\infty} \frac{d \omega}{2\,\pi\,i}e^{\omega\,Y}\,\tilde{n}\Lb \omega,r\Rb
\eeq
the equation has the form:
\beq \label{BFKLOM}
\omega\, \,\tilde{n}\Lb \omega,r\Rb\,\,=\,\,
- \,\,\int \frac{d^2 \vec{r}'}{2\,\pi}\,\omega_G\Lb \vec{r} - \vec{r}'\Rb \Bigg\{\tilde{n}\Lb \omega, r'\Rb\,\,-\,\,\h\tilde{n}\Lb \omega, r \Rb\Bigg\} \eeq
 This equation has been solved in Ref.\cite{LLS}.
 The main features of the solution can be summarized as follows:
\begin{itemize}
\item\quad The eigenvalues of  \eq{BFKLOM} are the same as for the
 BFKL equation with $m=0$ for $\omega\Lb \gamma \Rb\,\geq\,- \h \bas$,
 and can be parameterized in the following way:
 \bea \label{CHI}
\omega\Lb  \gamma\Rb\,\,&=&\,\,\bas\,\chi\Lb \gamma \Rb\,\,\,=\,\,\,\bas \Lb 2 \psi\Lb 1\Rb \,-\,\psi\Lb \gamma\Rb\,-\,\psi\Lb 1 - \gamma\Rb\Rb\\
&\,\xrightarrow{\gamma \to \h}& \,\omega_0\,\,+\,\,D\,\Lb \gamma - \h\Rb^2  \,\,+\,\,{\cal O}\Lb (\gamma - \h)^3\Rb\,\,=\,\,\bas 4 \ln 2  \,\,+\,\,\bas 14 \zeta\Lb 3\Rb \Lb \gamma - \h\Rb^2  \,\,+\,\,{\cal O}\Lb (\gamma - \h)^3\Rb\nn
 \eea
 where $\psi(z)$ is the Euler psi-function (see formulae {\bf 8.36} in Ref.
 \cite{RY}).
 \item\quad   The eigenfunctions $\phi_\gamma\Lb a; r\Rb$ have the  following
 behaviour:
  \beq \label{EIGEF}
\phi_\gamma\Lb a; r\Rb\,\,\rightarrow\,\,\left\{ \begin{array}{l l} r \,\rightarrow \infty\,\, & \,\,\, \propto ~~ e^{- a |r|}\\
  &\\
  r \,\rightarrow\,\, 0 &\,\,\, \propto\,~~r^{-2 + 2 \gamma} \end{array}\right.
  \eeq 
 \item\quad  In the   momentum representation for $\gamma \to \h$ the
 eigenfunctions $\phi\Lb \gamma, a; r\Rb$
 can be  written as
 
\beq \label{EIGEF1}
\phi_\gamma\Lb q, m\Rb\,\,=\,\,\frac{1}{\sqrt{q^2\, + \,4\,m^2}}\Lb \frac{\sqrt{q^2\, +\, 4\,m^2} + \sqrt{q^2}}{\sqrt{q^2\, + \,4\,m^2} - \sqrt{q^2}}\Rb^{\gamma - \h}
\eeq
  \item\quad  \eq{EIGEF1} means that the maximal intercepts
 $\omega\Lb \gamma\Rb$ reaches  the value  $\,4 \,\ln 2\,\bas$   at $\gamma
 = \h$,  as for  massless BFKL,  and  $a = 2\,m$.
  
    \end{itemize} 
 Expanding $\tilde{n}\Lb \omega,r\Rb$ in a series of the eigenfunctions
 $\phi_\gamma\Lb a; r\Rb$: viz.
 \beq \label{NTISER}
 \tilde{n}\Lb \omega,r\Rb\,\,=\,\,\int^{\epsilon\,+\,i\,\infty}_{ \epsilon\,-\,i\,\infty}\frac{d \gamma}{2\,\pi\,i}\,\phi_\gamma\Lb a; r\Rb \,n_{\rm in}(\gamma),
 \eeq
 where $n_{\rm in}(\gamma)$  is determined by the initial conditions,
 we obtain the solution to \eq{BFKLOM} in the form:
 \beq \label{SOLM2}
  \tilde{N}\Lb r, Y \Rb\,\,=\,\,\int^{\epsilon\,+\,i\,\infty}_{\epsilon\,-\,i\,\infty} \frac{d \omega}{2\,\pi\,i}e^{\omega\,Y}\, \int^{\epsilon\,+\,i\,\infty}_{ \epsilon\,-\,i\,\infty}\frac{d \gamma}{2\,\pi\,i}\,\phi_\gamma\Lb a; r\Rb \,\frac{1}{ \omega\,\,-\,\,\omega\Lb \gamma\Rb}\,n_{\rm in}(\gamma),
  \eeq
 where $\omega(\gamma)$ is given by \eq{CHI}.\
 
 The general solution for the inhomogeneous equation (see \eq{BFKLCRB})   
 has the form:
 
 \beq \label{SOLM3}
  N\Lb r,b;  Y \Rb\,\,=\,\, - \bas \omega_G\Lb 2 \,b\Rb \,\int^{\epsilon\,+\,i\,\infty}_{\epsilon\,-\,i\,\infty} \frac{d \omega}{2\,\pi\,i}e^{\omega\,Y}\, \int^{\epsilon\,+\,i\,\infty}_{ \epsilon\,-\,i\,\infty}\frac{d \gamma}{2\,\pi\,i}\,\phi_\gamma\Lb a; r\Rb \,\frac{1}{\Lb  \omega\,\,-\,\,\omega\Lb \gamma\Rb\Rb^2}\,n^0_{\rm in}(\gamma) \,\,+\,\,   \tilde{N}\Lb r, Y\Rb\,B\Lb b \Rb 
 \eeq
In \eq{SOLM3}, we used that $N\Lb r', b=0 , Y\Rb$ is equal to

 \beq \label{SOLM4}
N\Lb r, b=0 , Y\Rb\,\,=\,\,\int^{\epsilon\,+\,i\,\infty}_{\epsilon\,-\,i\,\infty} \frac{d \omega}{2\,\pi\,i}\,e^{\omega\,Y}\, \int^{\epsilon\,+\,i\,\infty}_{ \epsilon\,-\,i\,\infty}\frac{d \gamma}{2\,\pi\,i}\,\phi_\gamma\Lb a; r\Rb \,\frac{1}{ \omega\,\,-\,\,\omega\Lb \gamma\Rb}\,n^0_{\rm in}(\gamma)   
\eeq
where $ n^0_{\rm in}(\gamma)$ is determined by the initial
 condition: $N\Lb r, b=0 , Y = 0\Rb\,\,=\,\,{\cal N}\Lb r\Rb$.

The last term in \eq{SOLM4} is the solution to the homogeneous
  equation,  in which the function $B\Lb b \Rb$ is given by the initial 
condition.

\eq{SOLM3} leads to a scattering amplitude that decreases
 as $e^{ - 4\,m\,b}$.   Certainly such behaviour at large $b$,  
restores
 the Froissart theorem.

 
  \subsection{The size of the Froissart disc}
 
In the CGC approach, the scattering amplitude reaches the black disc limit
 $N\Lb r,b;Y\Rb\,\to\,1$ in the kinematic region:
 $ r^2\,Q^2_s(Y,b) \,\geq\,1$. Hence, we can find the size of the 
Froissart
 disc $R$,   from the equation:
\beq \label{SFD1}
r^2\,\,Q_s^2\Lb Y, R\Rb\,\,=\,\,1
\eeq

It is well known\cite{KOLEB,GLR,BALE,LETU, IIML,MUT}, that we don't  need
 to know the exact structure of the non-linear corrections to  find the
 saturation scale. We  only need  to solve the linear BFKL equation, and
 determine the line on which the scattering amplitude is constant.

The saturation  momentum $Q_s$ increases with energy and, therefore,
  small $ r\,\,\sim\,\,1/Q_s\,  \,\leq\,\,1/m$ 
 contribute to \eq{SFD1}. In this 
kinematic
 region we can use the eigenfunction 
$ \phi_\gamma\Lb a; r\Rb\,\, \propto\,~~r^{-2 + 2 \gamma} $  and 
 \eq{SOLM3} takes the form

 \bea \label{SFD2}
 && N\Lb r,b;  Y \Rb\,\,=\\
 &&\,\, e^{-4\,m,b} \,Y \,\int^{\epsilon\,+\,i\,\infty}_{\epsilon\,-\,i\,\infty} \frac{d \omega}{2\,\pi\,i} \int^{\epsilon\,+\,i\,\infty}_{ \epsilon\,-\,i\,\infty}\frac{d \gamma}{2\,\pi\,i}\,\,e^{\omega\Lb \gamma\Rb\,Y\,\,+\,\,(\gamma - 1) \xi}\,\,\,n^0_{\rm in}(\gamma) \,\,\,\equiv\,\,\,Y \,\int^{\epsilon\,+\,i\,\infty}_{\epsilon\,-\,i\,\infty} \frac{d \omega}{2\,\pi\,i} \int^{\epsilon\,+\,i\,\infty}_{ \epsilon\,-\,i\,\infty}\frac{d \gamma}{2\,\pi\,i}\,\,e^{\Psi\Lb Y; \xi, b; \gamma\Rb}\,\,\,n^0_{\rm in}(\gamma) \nn 
  \eea
  with $\xi\,\,=\,\,\ln\Lb r^2 \Lambda^2_{\rm QCD}\Rb$.

Using the method of steepest descent we can find the value of
 $\gamma_{cr}$ from the following two equations:

  \begin{subequations}  
  \bea 
  \mbox{Equation for saddle point}~~ \gamma_{\rm SP}: & \frac{\partial \,
\Psi\Lb Y; \xi, b; \gamma\Rb}{\partial\,\gamma}\,=\,0; & 
 \frac{d\,\omega(\gamma_{\rm SP})}{ d \gamma}\,Y\,\,+\,\,\xi
 \,\,=\,\,0;\label{SP1}\\
   \mbox{Equation for constant amplitude }\,:&\Psi\Lb Y; \xi,
 b; \gamma\Rb,\,=\,\,0;&\omega(\gamma_{\rm SP})\,Y\,\,+\,\,\Lb
 \gamma_{\rm SP} \,-\,1\Rb \xi\,\,-\,\,4\,m\,b\,=\,\,0;\label{SP2}
   \eea
     \end{subequations} 
     
\begin{figure}
\centering
\begin{tabular}{c c c}
 \includegraphics[width=8cm]{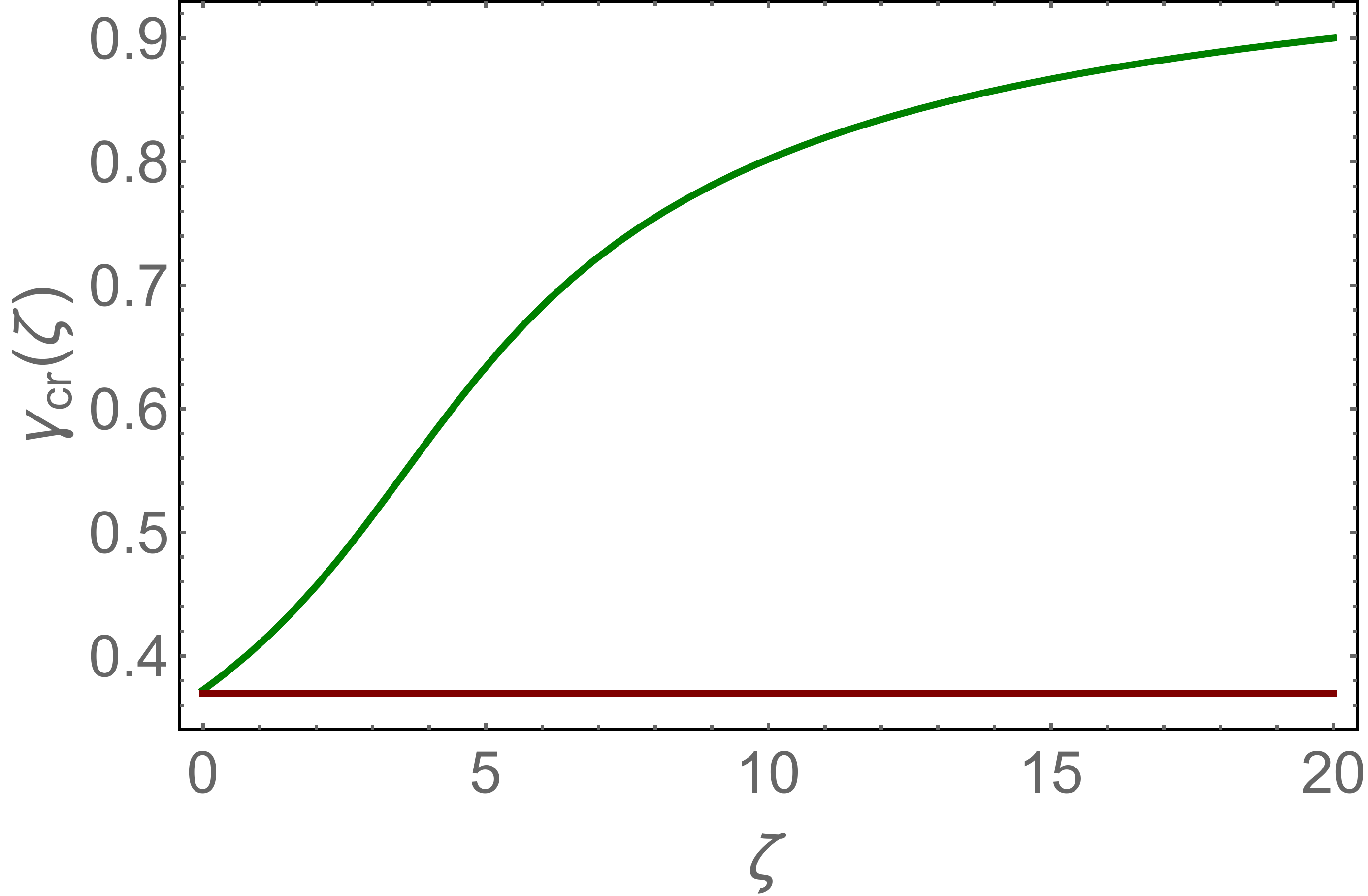} & ~~~& \includegraphics[width=8cm]{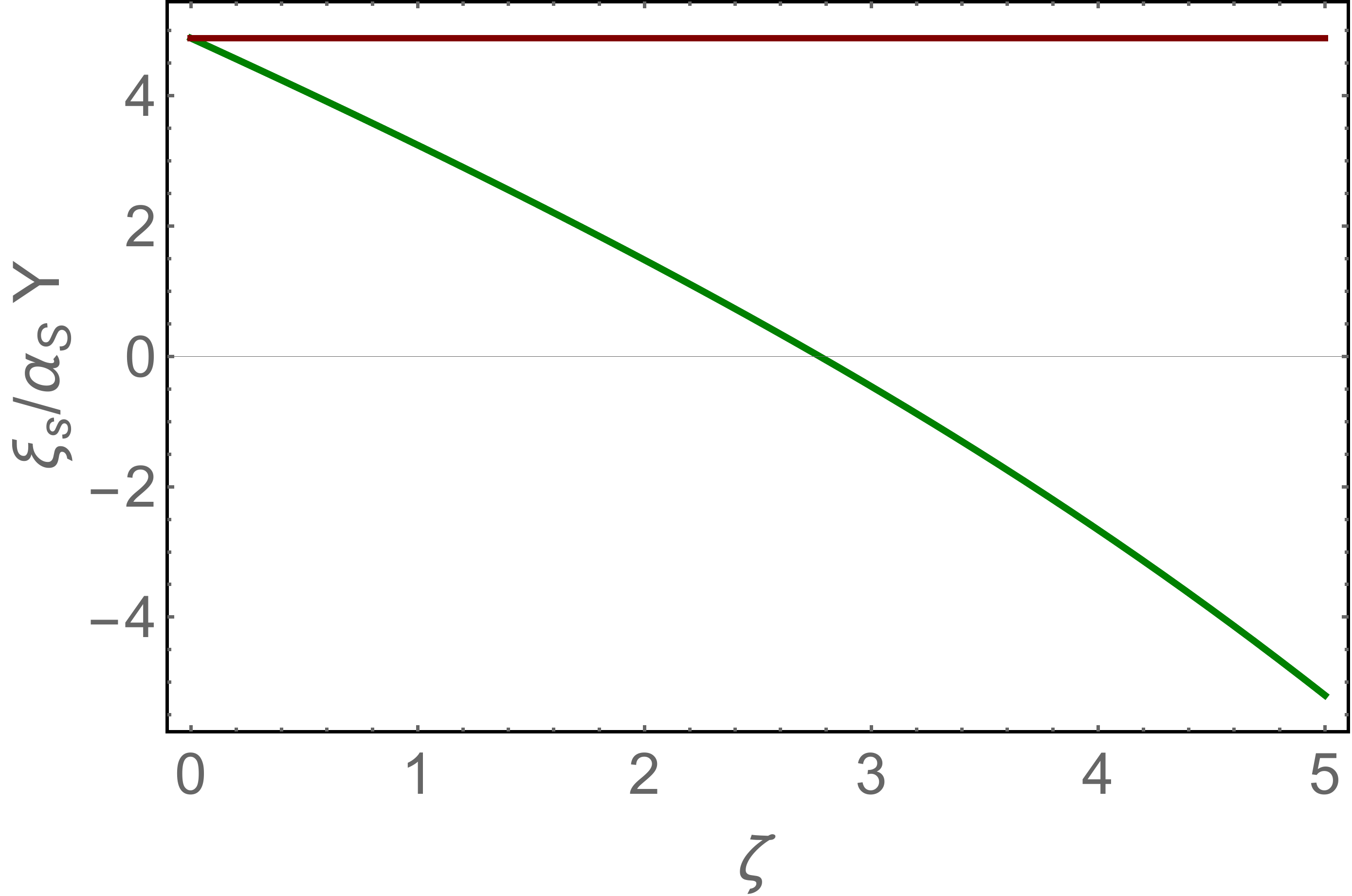}\\
 \fig{gacr}-a & & \fig{gacr}-b\\
 \end{tabular}
   \caption{\fig{gacr}-a:  $\gamma_{cr}$ versus $\zeta$ from  \eq{GACR}.
 The red line corresponds to  $\gamma_{cr} =0.37$, which is the solution 
of
 \eq{GACR} at $b=0$.\fig{gacr}-b: $\xi_s\Big{/}\bas Y$ versus $\zeta$,
 where $\xi_s\,\,=\,\,\ln\Lb Q^2_s\Lb Y, \zeta\Rb/ Q^2_s\Lb Y = 0,
 \zeta = 0\Rb\Rb$. The red line shows  $\xi_s\Big{/}\bas Y$ at $\zeta = 
0$.}    
\label{gacr}
\end{figure}

     Solving \eq{SP1} and \eq{SP2} we obtain an equation for
 $\gamma_{SP} = \gamma_{cr}$, which has  the form:
     \beq \label{GACR}  
     \chi\Lb \gamma_{cr}\Rb \,+\,\Lb 1 \,-\,\gamma_{cr}\Rb\,\frac{d\,\chi(\gamma_{cr})}{ d \gamma}  \,\,=\,\,
     \zeta ~~~\mbox{where}~~~\zeta \,=\,\frac{4\,m\,b}{\bas\,Y}
       \eeq 
      The  solution to \eq{GACR} is shown in \fig{gacr}-a. One can see
 that the value of $\gamma_{cr}$ depends on the value of $\zeta$.

        From \eq{SP1} we can calculate $\xi_s$, which is equal to 
        \beq \label{XIS}
       \xi_s\,\,=\,\,\ln\Lb Q^2_s\Lb Y, \zeta\Rb/ Q^2_s\Lb Y = 0,
 \zeta = 0\Rb\Rb\,\,=\,\,-\, \frac{d\,\chi(\gamma_{\rm SP})}
{ d \gamma}\Big{|}_{\gamma_{\rm SP}\,=\,\gamma_{cr}\Lb \zeta\Rb}\,\bas\, Y
       \eeq
       In \fig{gacr}-b we plot  the value  $\xi_s\Big{/}\bas Y$ as a 
function of $\zeta$. For $\zeta\,>\,2.8$, the
 saturation momentum starts to decrease as function of $Y$. In 
the vicinity of the saturation scale the scattering amplitude 
has the following form\cite{MUT}:
       \beq \label{VQS}
       N\Lb r, b; Y\Rb\,\,=\,\,N_0 \Lb r^2\,Q^2_s\Lb  Y, \zeta\Rb\Rb^{1\,-\,\gamma_{cr}\Lb \zeta\Rb}
       \eeq
       where $N_0$ is a constant smaller than 1.

       The radius of the Froissart disc ($R$) can be found from the condition:
       \beq \label{RFD}
         N\Lb r, R; Y\Rb \,\,=\,\,f
         \eeq

         where $f$ is a constant ( $f\,<\,1$ ). Introducing a new
 variable $z\,=\,\ln\Lb r^2\,Q^2_s\Lb Y, \zeta\Rb\Rb$,  \eq{RFD} can
 be re-written as
             \beq \label{RFD}
      \Lb 1 - \gamma_{cr}\Lb \zeta_R \Rb \Rb\,z \,\,=\,\, \Lb 1 - \gamma_{cr}\Rb\, \Lb   \xi  + \xi_s \Rb\,\,=\,\, \Lb 1 - \gamma_{cr}\Rb\,\Lb\xi \,-\,\, \frac{d\,\chi(\gamma_{\rm SP})}{ d \gamma}\Big{|}_{\gamma_{\rm SP}\,=\,\gamma_{cr}\Lb \zeta_R\Rb, \zeta= \zeta_R}\,\bas\, Y   \Rb   
      =\,\,\tilde{f}
         \eeq                  
       where $\tilde{f} \,=\,\ln\Lb\frac{f}{N_0}\Rb$. We re-write
 \eq{RFD}  as follows:
              \beq \label{RFD1} 
       \eta\,\,-\,\,\frac{d\,\chi(\gamma_{\rm SP})}{
 d \gamma}\Big{|}_{\gamma_{\rm SP}\,=\,\gamma_{cr}\Lb
 \zeta_R\Rb, \zeta= \zeta_R}\,\,=\,\,\frac{\tilde{f}}
{\bas Y}\,\,\xrightarrow{\bas Y \,\gg\,1} 0;~~~~~~~~~
\eta\,\,=\,\,\frac{\xi}{\bas Y}
       \eeq

\begin{figure}
\centering
\begin{tabular}{c c c}
 \includegraphics[width=8cm]{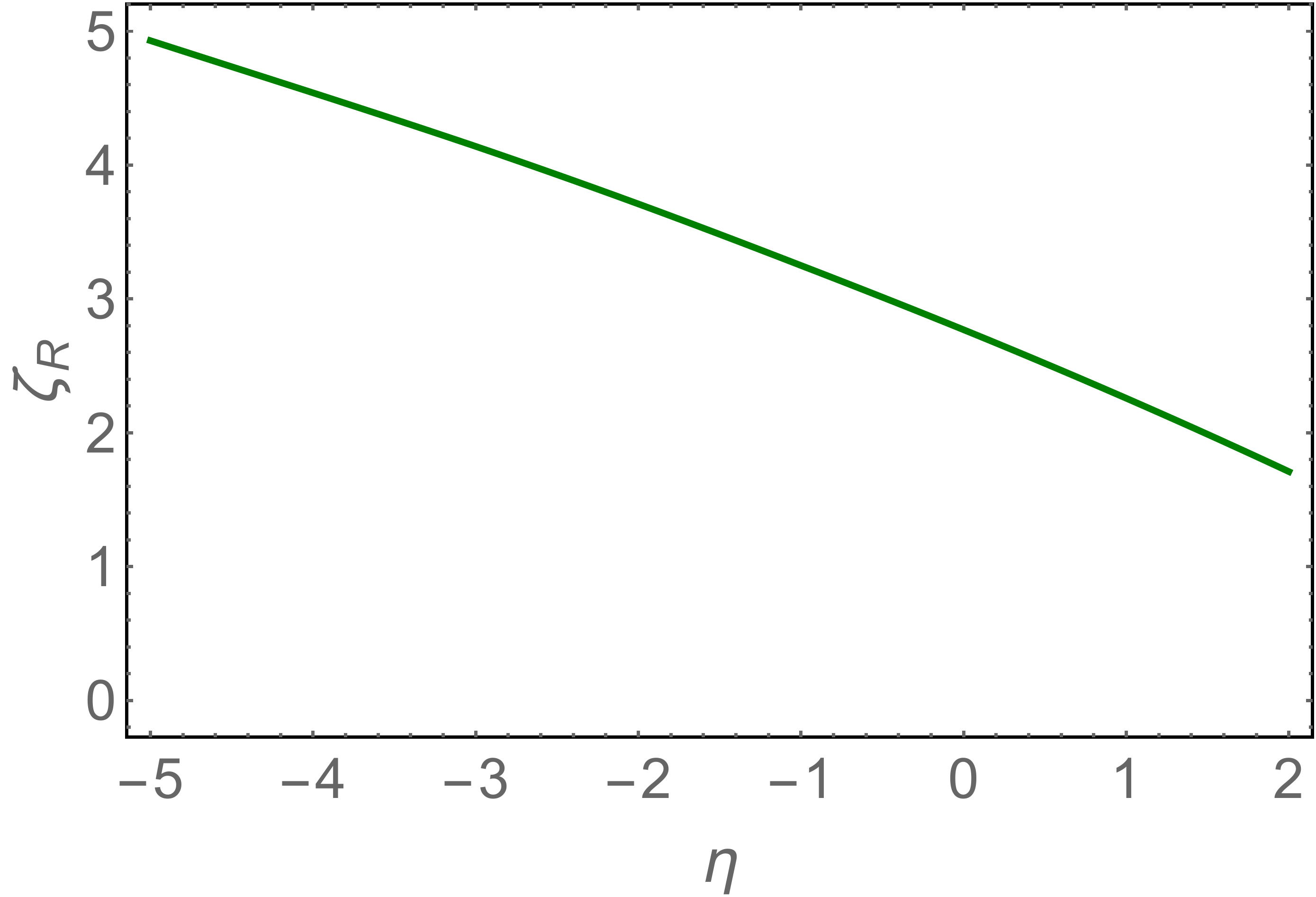} & ~~~& \includegraphics[width=8cm]{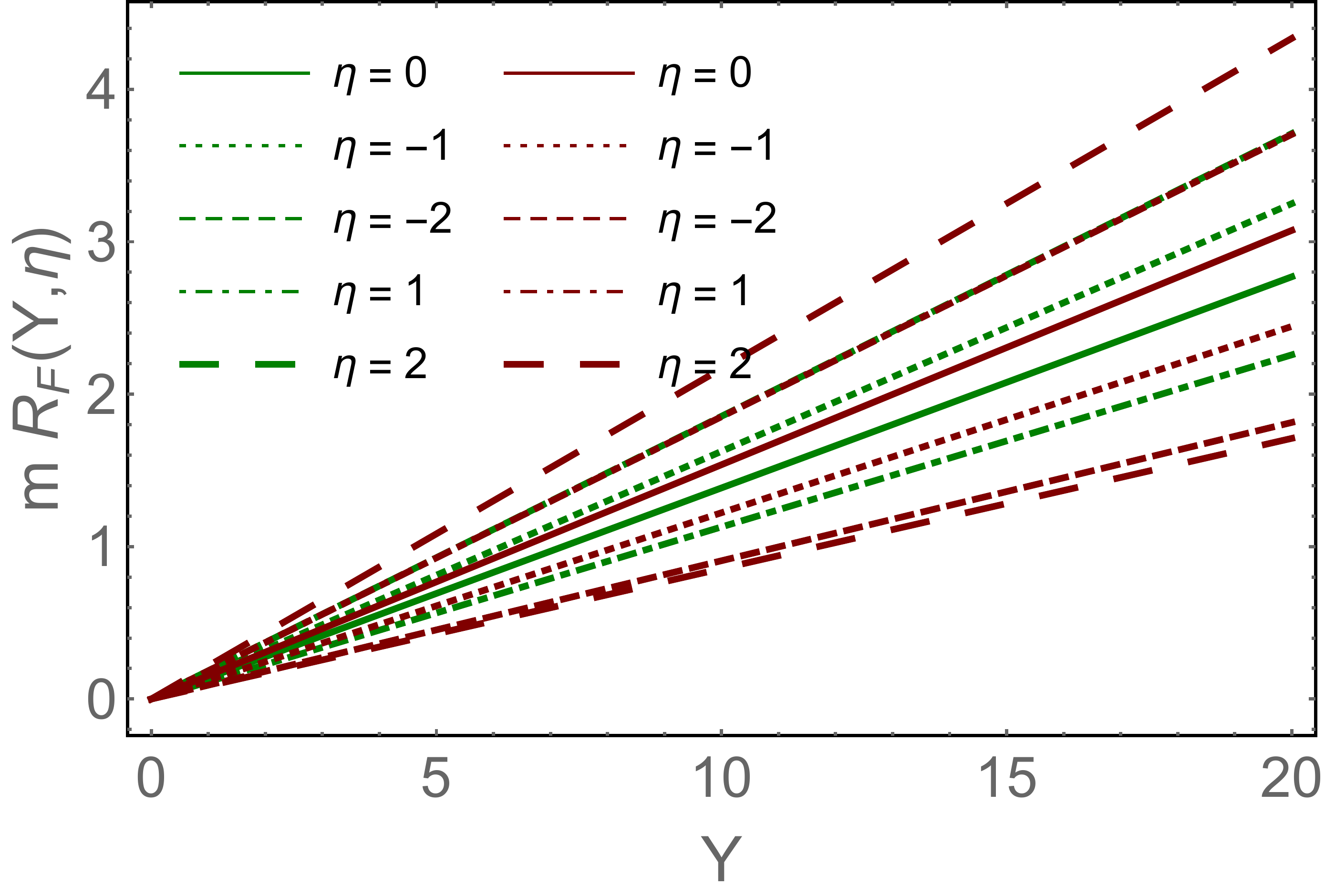}\\
 \fig{rf}-a & & \fig{rf}-b\\
 \end{tabular}
   \caption{\fig{rf}-a:  $\zeta_R$ versus $\eta$ from \eq{RFD1}. 
   \fig{rf}-b: The radius of the Froissart disc versus $Y$ at different
 values of $\eta$.The red lines show the estimates from \eq{B0} for
 the radius of the Froissart disc. The value of $\bas$ is taken
 $\bas = 0.2$.}    
\label{rf}
\end{figure}
 In \fig{rf} we plot the solution to \eq{RFD1} as a function of
  $\eta$, while in \fig{rf}-b we see the dependence of the radius
 of the Froissart disc on $Y$. Note that the radius increases
  as ${\rm Const}\, Y$, but the value of 
  ${\rm Const} $ depends on $\xi$($\eta$).              
     
\subsection{Discussion}

       
  Hence, we can conclude that in non-abelian gauge theories with the
 Higgs mechanism for   mass generation,    
   in the CGC approach, we obtain  a Froissart disc with the radius $R 
\,\propto\,Y$,
 with a  coefficient of proportionality, which depends on the size
 of colliding dipole.
   
     
\subsubsection{Restoration of the Froissart  theorem}

     It is easy to demonstrate the restoration of the Froissart
 theorem\cite{FROI} for this approach.   Using the unitarity
 constraints that  $N\Lb r, b; Y\Rb \,\,\leq\,\,1$ , we can 
find the bound for the total cross section (see for example 
appendix 2.2 of Ref.\cite{KOLEB}):
    \beq \label{DIS1}
\sigma_{tot}\,\,=\,\,2 \int N\Lb r, b; Y\Rb d^2 b\,\,<\,\,\underbrace{2 \int^{b_0} d^2 b}_{\mbox{unitarity constrants}}\,\,+\,\,\int_{b_0}d^2 b\, \,N\Lb r, b; Y\Rb 
\eeq   
  We estimate the value of $b_0$, using the following equation

\beq \label{DIS2}
 N\Lb  r , b_0;  Y \Rb\,\,=\,\,f\,<1
 \eeq
Plugging in \eq{DIS2} the solution of the BFKL equation in the
 form: $\tilde{N}\Lb r; Y\Rb\,\exp\Lb - \,4\,m\,b\Rb$ (see 
   \eq{SOLM3}) we obtain
   \beq \label{DIS3}
    \tilde{N}\Lb  r ;  Y \Rb\,e^{ - 4\,m\,b_0}\,\,=\,\, \underbrace{N_0 \Lb r^2 Q_s(Y)\Rb^{1 - \gamma_{cr}}}_{\mbox{vicinity of the saturation momentum}}\,e^{ - 4\,m\,b_0} \,\,=\,\,   f\,<1       
      \eeq
      where $\gamma_{cr}\,\,\approx\,\,0.37$ is the solution to
 \eq{SP1} and \eq{SP2} at $m\,=\,0$. From \eq{DIS3} one can see that
      \beq \label{B0}
      b_0\,\,=\,\,\frac{1}{4\,m}\Big((1 - \gamma_{cr})
      \,z\,\,-\,\,\ln\frac{f}{N_0}\Big)
      \eeq
      where $z\,\,=\,\,\ln\Lb r^2 Q_s(Y)\Rb\,\,=\,\,\xi\,\,- \,
\bas \frac{ d \chi\Lb \gamma\Rb}{d\,\gamma}\Big{|}_{\gamma\,=\,
\gamma_{cr}}\,Y$.
      The dependence of the radius of the Froissart disc given
 by \eq{B0}, is shown in \fig{rf}-b by red lines. One can see
 that in spite of the same proportionality to $Y$, the value of the
 coefficients are quite different.

     Plugging \eq{B0} into \eq{DIS1}  one can see that
     \beq \label{DIS4}
     \sigma_{tot}\,\,\leq \,\,2 \pi\,b^2_0\,\,=\,\, 2\,\pi \Lb  \frac{1}{4\,m}\Big(\frac{1}{1 - \gamma_{cr}}\,z\,\,-\,\,\ln\frac{f}{N_0}\Big)\Rb^2
   \eeq
   
    For $\xi=0$ \eq{DIS4} gives $  \sigma_{tot}\,\,\leq\,\,  
 2\,\pi \Lb\bas \, \chi(\gamma_{cr}\,Y\Rb^2$ and, therefore, 
    leads to the Froissart theorem.

     
\subsubsection{More about eigenfunctions - a recap}

  To learn more about the behaviour of the eigenfunction at
 large distances we follow Ref.\cite{LLS} and consider            
    the BFKL equation (see \eq{KEREMM} and \eq{GLTR})  at $Q_T = 0$. 
   It has the form:
    \bea \label{DIS5}
\hspace{-0.5cm}&&E \phi\Lb \kappa\Rb\,\,=\\
\hspace{-0.5cm}&&\,\,\underbrace{\frac{\kappa +1}{\sqrt{\kappa}\sqrt{\kappa + 4  }}\ln \frac{\sqrt{\kappa + 4 } + \sqrt{\kappa}}{\sqrt{\kappa + 4 } - \sqrt{\kappa}} \phi\Lb \kappa\Rb}_{{\rm kinetic\, energy\, term}}\,\,-\,\,\underbrace{\int^{\infty}_{0}\,\frac{d \kappa' \phi\Lb \kappa'\Rb}{\sqrt{( \kappa - \kappa')^2\,+\,2 (\kappa + \kappa') + 1}}}_{\rm potential\, energy\, term}\,\,+\,\,\underbrace{\frac{N^2_c + 1}{2 N^2_c}\frac{1}{\kappa + 1}\int^{\infty}_0 \frac{\phi\Lb \kappa'\Rb \,d \kappa'}{\kappa' + 1}}_{{\rm contact\, term}}\nn
\eea
         
 In \eq{DIS5} we introduce the following notations:
 \beq \label{VAR}
\kappa\,=\,\frac{q^2}{m^2};\,\,\,\,\,\,\,\,\,\,\,\,\,\,\kappa'\,=\,\frac{q'^2}{m^2};\,\,\,\,\,\,\,\,\,\,\,\,\,E\,=\,- \frac{\omega}{\bas};\,\,\,\,\,\,\,\,\,\,\,\,\, \bas\,=\,\frac{\as N_c}{\pi}
\eeq 
   Re-writing \eq{DIS5} in the coordinate representation, we can see that
 it takes the form:          
\beq \label{H}
E\phi\Lb r \Rb\,\,=\,\,{\cal H}\,\phi\Lb r \Rb
\eeq
with
\beq\label{H1}
{\cal H}\,=\,\frac{\kappa^2 + 1}{|\kappa|\sqrt{\kappa^2 + 4 
 }}\ln \frac{\sqrt{\kappa^2 + 4 } + |\kappa|}{\sqrt{\kappa^2
 + 4 } - |\kappa|}\,\,-\,\,2 K_0\Lb |r| m \Rb\,+\,\frac{N^2_c
 + 1}{2\,N^2_c} \hat{P}\,\,=\,\,T\Lb \kappa \Rb \,\,+\,\,V\Lb r
 \Rb \,\,+\,\,\frac{N^2_c + 1}{2\,N^2_c} \hat{P}\eeq
where $\hat{P}$ is  a shorthand notation for the projector onto the state
 $\sim m^2/(p^2+m^2)$
\beq \label{P}
\hat{P} \,\phi\Lb \kappa \Rb\,=\,\frac{1}{\kappa^2 + 1}\int \frac{d^2 \kappa'}{\pi} \frac{\phi\Lb \kappa'\Rb}{\kappa'^2\,+\,1}
\eeq
where $\kappa \,=\,-i \nabla_\perp$.

At large distances ($ r \to \infty$) the potential energy in 
the Hamiltonian
 $\Lb V\Lb r \Rb = - 2 K_0\Lb r m \Rb\Rb$ is exponentially small, the 
contribution from the projector $\hat{P}$ in \eq{H} is proportional to
 $1/(\kappa^2 + 1)$ and is also exponentially suppressed, so the only
 relevant term in the Hamiltonian is the kinetic energy
\beq \label{KINE}
 E\Lb \kappa\Rb\,\,=\,\,T\Lb \kappa \Rb\,\,=\,\,\frac{\kappa^2 + 1}{|\kappa|\sqrt{\kappa^2 + 4  }}\ln \frac{\sqrt{\kappa^2 + 4 } + |\kappa|}{\sqrt{\kappa^2 + 4 } - |\kappa|},
 \eeq
 for which the eigenfunctions have a form
\beq \label{PLWA}
 \phi\Lb \vec{r}\Rb\,\,\sim\,\, e^{ i \sqrt{\kappa^2} r}, ~~\kappa^2\,\,>\,\,0;\quad \phi\Lb \vec{r}\Rb\,\,\sim\,\, e^{ - \sqrt{-\kappa^2} r},~~ \kappa^2\,<\,0.
 \eeq
The point $\kappa \,=\,0$ is special, since it separates two different
 behaviours at large  $r$. This point corresponds to energy $E=\h$ or
 $\omega=\omega_0\equiv -\h\bas$. As   is shown in Ref.\cite{LLS},
 there are
 qualitative changes in the shape of the wave functions near this
 point. From the structure of the kinetic energy term~(\ref{KINE})
 we can see that the energy $E$ is positive ($\omega<0$) for
 $\kappa^2>0$, however for $-4\,<\,\kappa^2\,<\,0$, the energy
 may  have any value from $-\infty$ up to $\h$. This means that
 for $\kappa^2\,<\,0$ we have a discrete spectrum with two
 conditions  shown in \eq{EIGEF}.  Hence, the exponential
 decrease of the eigenfunction is intimately related to the
 behaviour of the reggeization term in the BFKL equation, and
 it stems from the region, where  $\omega\Lb q\Rb$ is positive.

     \begin{figure}[ht]
     \begin{center}
     \includegraphics[width=8cm]{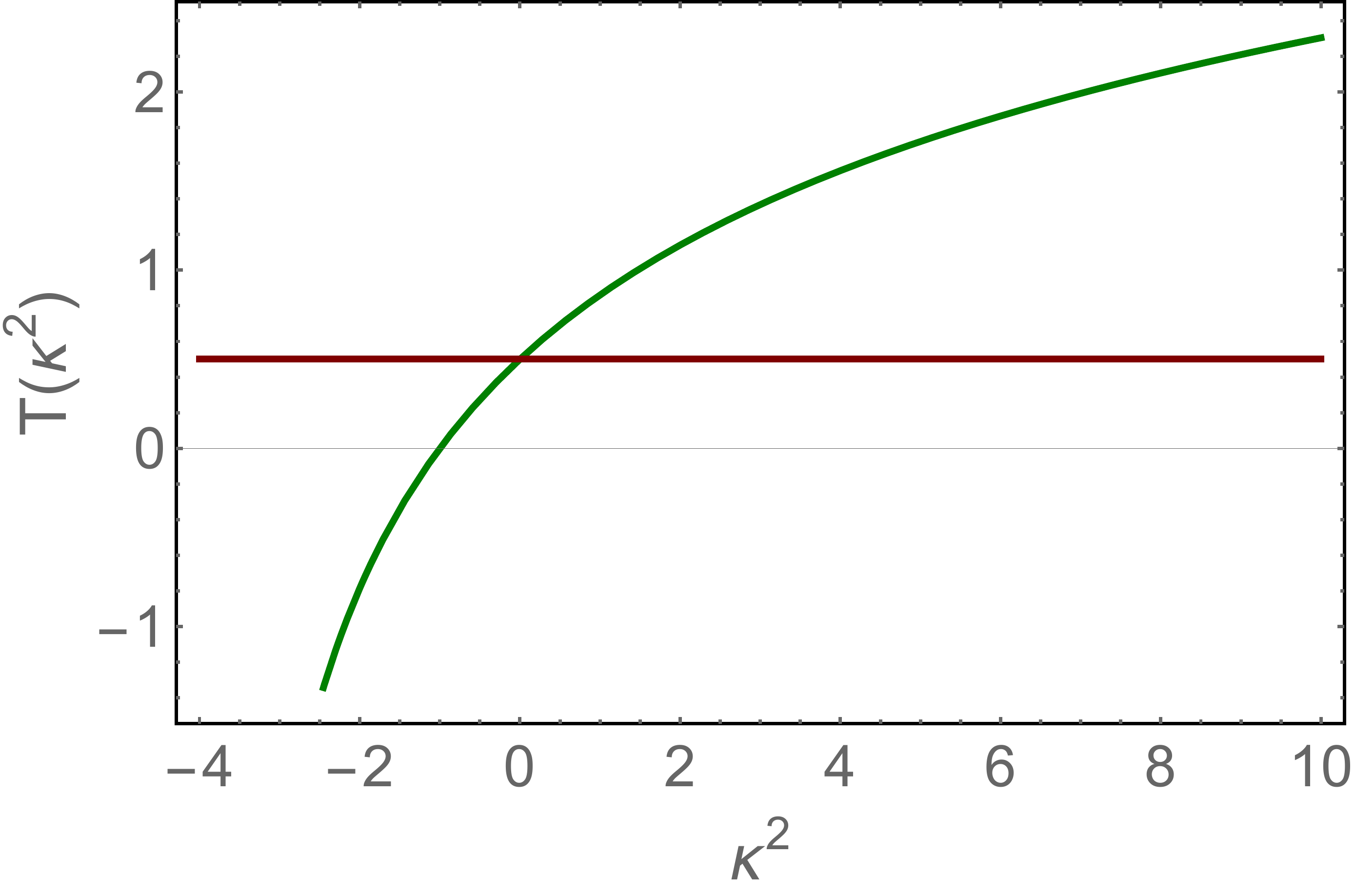} 
     \end{center}    
      \caption{ The dependence of the kinetic energy (see \protect\eq{KINE})
 versus $\kappa^2$ . The red line corrsponds to $T\Lb \kappa=0\Rb $.}
\label{t}
   \end{figure}

    The large $b$ dependence is determined by the singularities of this term
 which in  turn, corresponds   to the singularities of the gluon
 propagator.  In this model it is a pole at  the Higgs mass. Actually,
 the scattering amplitude at large $b$
 $ N\Lb r, b; Y\Rb \,\xrightarrow{r\,m\,\gg\,1} \,\exp\Lb - \,4\,m\,b\Rb $,
 where   $q^2  = -  4\,m^2$, is the singularity of the gluon reggeization 
in
 the momentum space (see \eq{GLTR}). Hence, our next step will be to
 understand the singularities of the gluon propagator in QCD. Certainly,
 they have a non-perturbative origin, and  we have to rely on 
 a  non-perturbative approach, which is in an embryonic stage at the
 moment. The only reliable information comes from lattice QCD\cite{LAT},
 which we will discuss in the next section.

     
\section{Gribov - Zwanziger confinement and  the BFKL equation}
Among numerous approaches to confinement, the one proposed by Gribov, 
\cite{GRI0,GRI1,GRI2,GRI3,GRI4,GRREV,DOKH,KHLE} has special
 advantages ,which makes it most suitable 
for discussion of the BFKL equation in the framework of this hypothese.
 First, it is based on the existence of Gribov copies\cite{GRI0} - multiple
 solutions of the gauge-fixing conditions, which are the principle 
properties of  non-perturbative QCD. Second, the main ingredient
 is the modified gluon propagator, which can be easily included
 in the 
BFKL-type of equations. Third, in Ref.\cite{KHLE}(see also ref.\cite{FDGS})
 it  is demonstrated that  the Gribov gluon propagator originates 
naturally
  from the topological structure of non-perturbative QCD 
 in the form:

\beq \label{GLPR}
G\Lb q\Rb\,\,=\,\,\frac{1}{q^2\,\,+\,\,\frac{\chi_{\rm top}}{q^2}}\,\,\,=\,\,\frac{q^2}{q^4\,\,+\,\,\mu^4}\,\,=\,\,\h\Bigg( \frac{1}{q^2 \,+\,i\,\mu^2} \,\,+\,\, \frac{1}{q^2 \,-\,i\,\mu^2}\Bigg)
\eeq
where $\chi_{\rm top}\,\,=\,\,\mu^4 $ is the topological susceptibility 
of QCD, which is related to  the $\eta'$ mass by the Witten-Veneziano
 relation\cite{VEN,WIT}. This allows us to obtain the principal
 non-perturbative  dimensional scale, directly from the experimental data.

\begin{boldmath}
\subsection{ The gluon propagator. }
\end{boldmath}

     As we have discussed above, to find  the large impact parameter
 behaviour, we need to know the gluon reggeization contribution in  
coordinate space. However, before calculating it, we evaluate the 
behaviour of the gluon propagator.  As we can see from \fig{eq}, the 
gluon reggeization term comes from the exchange of gluons at high
 energy. It is known (see Ref.\cite{KOLEB}) that $t$-channel gluons
 in the BFKL equation depend only on transverse momenta of the
 gluons. Hence, we need to calculate 
   the following integral in coordinate space:
      
    \beq \label{GLPR1}
    G\Lb r  \Rb\,\,\,=\,\,\,\int \frac{d^2 q_T}{(2 \,\pi)^2} e^{i \vec{r}\cdot\vec{q}_T } \,G\Lb q_T\Rb
    \eeq
    Plugging in \eq{GLPR1} \eq{GLPR} we obtain
      \beq \label{GLPR2}
    G\Lb r  \Rb\,\,\,=\,\,\,\int \frac{d^2 q_T}{(2 \,\pi)^2} e^{i \vec{r}\cdot\vec{q}_T } \,\frac{q^2_T}{q^4_T\,\,+\,\,\mu^2}\,\,=\,   
   \frac{1}{4} G_{0,4}^{3,0}\left(\frac{r^4 \mu ^4}{256}|
\begin{array}{c}
 0,0,\frac{1}{2},\frac{1}{2} \\
\end{array}
\right)   
    \eeq     
    
    Where $G_{0,4}^{3,0}$ is the  Meijer's  G-Function (see formulae
 {\bf 9.3} given in  Ref.\cite{RY}). 
    
     \beq \label{GLPR3}
 G\Lb r  \Rb\,\,\rightarrow\,\,\left\{ \begin{array}{l l} r \,\rightarrow \infty\,\, & \,\,\, \propto ~~ e^{ - \frac{\mu\,r}{\sqrt{2}}}\,\cos\Lb \frac{\mu\,r}{\sqrt{2}}\,\,\frac{\pi}{8}\Rb\\
  &\\
  r \,\rightarrow\,\, 0 &\,\,\, \propto\,~~- \ln\Lb \frac{\mu\,r}{\sqrt{2}}\Rb\,-\,C \end{array}\right.
  \eeq          
    where $C$ denotes  the Euler constant.   
       
       Hence, we see that at large values of $r$ the gluon propagator
 decreases exponentially, giving us  hope, that  Gribov's confinement
 will lead to a scattering amplitude, that will be exponentially
 small at long distances.
  
\begin{boldmath}
\subsection{ The gluon trajectory. }
\end{boldmath}

  The general expression for the gluon trajectory has the following 
form\cite{FKL}:
  
  \beq \label{GGLTR}
  \omega_G\Lb q\Rb\,\,=\,\,G^{-1}\Lb q\Rb\Sigma\Lb q\Rb ~~~\mbox{where}
~~\Sigma\Lb q\Rb\,\,=\,\,\int \frac{d^2 q'}{4\,\pi} G\Lb \vec{q}'\Rb\,G\Lb
 \vec{q} - \vec{q}'\Rb\eeq
   
   Before making  estimates with the gluon propagator of \eq{GLPR}, we need
 to mention, that the lattice calculation of the gluon propagator leads to
 $G\Lb 0\Rb\,\neq\,0$( see Refs.\cite{DOS,DOV,CDMV} and references therein),
 in explicit contradiction with \eq{GLTR}. However,  in
 Ref.\cite{DSVV,DGSVV,DSV} it is proven that Gribov's copies generate the 
gluon
 propagator in a more general form:
   \beq \label{GGLPR1}
   G\Lb q\Rb \,\,=\,\,\frac{q^2\,\,+\,\,M^2_0}{\Lb q^2 \,+\,M^2\Rb^2\,\,+\,\,\mu^4}
   \eeq
  which leads to  $G\Lb 0\Rb\,\neq\,\,0$. 
 We consider this form as a parameterization of the sum of Gribov's 
propagators
 of \eq{GLPR},  with  different values of $\mu$. In particular, in
 Ref.\cite{DKEL} it was demonstrated  that the approach, suggested in
 Ref.\cite{KHLE} ,   leads to a gluon propagator of the following form:
   \beq \label{OURP}
   G\Lb q \Rb \,=\,\frac{1}{\sqrt{\pi}}\,\frac{1}{q^2}\,\int^{\infty}_0 d\zeta \,e^{- \zeta}\,\frac{\zeta^{-\h}}{1 \,+\,\zeta\,z} \,\,  =\,  \frac{\sqrt{\pi}}{\mu^2}\,e^{1/z}\,{\rm Erfc}\Lb \frac{1}{\sqrt{z}}\Rb
   \rightarrow\,\,\left\{ \begin{array}{l l} \,\,\,q^2\,\gg\,\mu^2 &  \propto\, \,\,1/q^2; \\
 & \\
  q^2\,\ll\,\mu^2  & \propto\, \,\sqrt{\pi}/\mu^2 ; \end{array}\right.
  \eeq 
   
where  $z \,=\,\frac{\mu^4}{q^4}$.

  As we have mentioned,
 at high energies $q$ is a two dimensional vector, which corresponds
 to transverse momentum  carried by the gluon. Introducing 
  \beq \label{GGLPR2}
  G^{\pm}\Lb q\Rb\,\,=\,\,\frac{1}{\Lb q^2\,\,+\,\,M^2\Rb\,\,\pm\,i\,\mu^2} 
  \eeq
  we can re-write \eq{GGLPR1} in the form:
   \bea \label{GGLPR3}
  &&G\Lb q\Rb\,\,=\\
  &&,\,\h \Big( G^{+}\Lb q\Rb \,\,+\,\,G^{-}\Lb q\Rb \Big) \,\,+\,\,\frac{M^2_0 - M^2}{2\,\mu^2\,i} \Big( G^{+}\Lb q\Rb \,\,-\,\,G^{-}\Lb q\Rb \Big)\,\,=\,\,\frac{1}{\mu^2}\Big( {\rm Re}\,G^{+}\Lb \kappa\Rb  \,\,+\,\,(M^2_0 - M^2)\, {\rm Im}\,G^{+}\Lb \kappa\Rb  \Big)\nn\\
  &&= \h\Bigg\{ (1\,\,+\,\,i\,\frac{M^2_0 - M^2}{\mu^2}) \,G^{+}\Lb q\Rb  + (1\,\,-\,\,i\,\frac{M^2_0 - M^2}{\mu^2}) \,G^{-}\Lb q\Rb  
  \Bigg\} = \h\Bigg\{ (1\,\,+\,\,i\,m_0) \,G^{+}\Lb \kappa\Rb \,\,+\,\,(1\,\,-\,\,i\,m_0) \,G^{-}\Lb \kappa\Rb  
  \Bigg\}  \nn   \eea  
   
    where we use notation   similar to  \eq{VAR}:
   \beq \label{VAR1}
\kappa\,=\,\frac{q^2}{\mu^2};\,\,\,\,\,\,\,\,\,\,\,\,\,\,\kappa'\,=\,\frac{q'^2}{\mu^2};\,\,\,\,\,\,\,\,\,\,\,\,\,E\,=\,- \frac{\omega}{\bas};\,\,\,\, \bas\,=\,\frac{\as N_c}{\pi}; \,\,\,m \,\,=\,\,\frac{M^2}{\mu^2}; \,\,\,m_0 \,\,=\,\,\frac{M^2_0\,-\,M^2}{\mu^2};\eeq 
    
    Plugging \eq{GGLPR3} into \eq{GGLTR} one can see that 
    \beq \label{GGLTR2}
    \Sigma\Lb \kappa\Rb\,\,=\,\,\int\frac{d^2 \kappa'}{4\,\pi} \sum_{j= \pm,\, l= \pm} a_{j l}\, G^{j}\Lb \vec{\kappa}'\Rb G^{l}\Lb \vec{\kappa}\,-\,\vec{\kappa}'\Rb
    \eeq
    where the coefficient $a_{j l}$ can be easily calculated from the
 decomposition of \eq{GGLPR3}. Each term of \eq{GGLTR2}  can 
be re-written in the form
    \bea \label{GGLTR3}
 &&   \Sigma_{1,2}\Lb \kappa\Rb\,\,=\,\,\int\frac{d^2 \kappa'}{4\,\pi} \, G^{1}\Lb \vec{\kappa}'\Rb G^{2}\Lb \vec{\kappa}\,-\,\vec{\kappa}'\Rb\,\,=\,\,\int\frac{d^2 \kappa'}{4\,\pi} \frac{1}{(\kappa'^2\,+\,m^2_1)\, ((\vec{\kappa}\,-\,\vec{\kappa}')^2\,+\,m^2_2)}\\
   && =\h \int^1_0 d \alpha\int\frac{d \kappa'^2}{\Lb ( \vec{\kappa}' - \alpha\,\vec{\kappa})^2 \,+\,\kappa\,\alpha (1 - \alpha) \,\,+\,m^2_1 \,-\,(m^2_2\,-\,m^2_1) \alpha\Rb^2}\,\,=\,\,\h \int^1_0 d \alpha \frac{1}{\,\kappa\,\alpha (1 - \alpha) \,+\,m^2_1 \,-\,(m^2_2\,-\,m^2_1) \alpha} \nn   
      \eea      
   where we have introduced the Feyman parameter $\alpha$ and 
$m^2_1=\,m\,\pm
 i $ and $ m^2_2\,=\,m\,\pm\, i $.
   
   Integrating over $\alpha$ we obtain:
   
   \beq \label{SIGMA}
   \Sigma_{1,2}\,=\,\frac{1}{\Delta}
   \displaystyle{\Big\{\ln\Big(\frac{-\kappa \,+ \,m^2_2\,-\,m^2_1\,
+\,\Delta}{\kappa \,+ \,m^2_2\,-\,m^2_1\,+\,\Delta} \Big)\,\,\,-\,
\,\ln\Big(\frac{\kappa \,- \,m^2_2\,+\,m^2_1\,+\,\Delta}{-\kappa \,-
 \,m^2_2\,+\,m^2_1\,+\,\Delta} \Big)} \Big\}
   \eeq
   where $\Delta \,\,=\,\,\sqrt{-\Lb \kappa + (m_1 - m_2)^2\Rb\,\Lb 
\kappa\,+\,(m_1\,+\,m_2)^2\Rb}$ and $\kappa$ is defined in \eq{VAR}.
   
   Cumbersome,   but simple calculations lead from \eq{SIGMA} to the
 expression for the gluon trajectory $T\Lb \kappa\Rb \,\,=\,\, 2\,
\Omega_G\Lb \kappa\Rb$ (see appendix A). \fig{sigma}-a shows the
 resulting $T(\kappa)$ as function of $\kappa$ for different
 values of $m$ and $m_0$. 
   
\begin{figure}[ht]
     \begin{center}
       \begin{tabular}{c c c}
     \includegraphics[width=8.5cm]{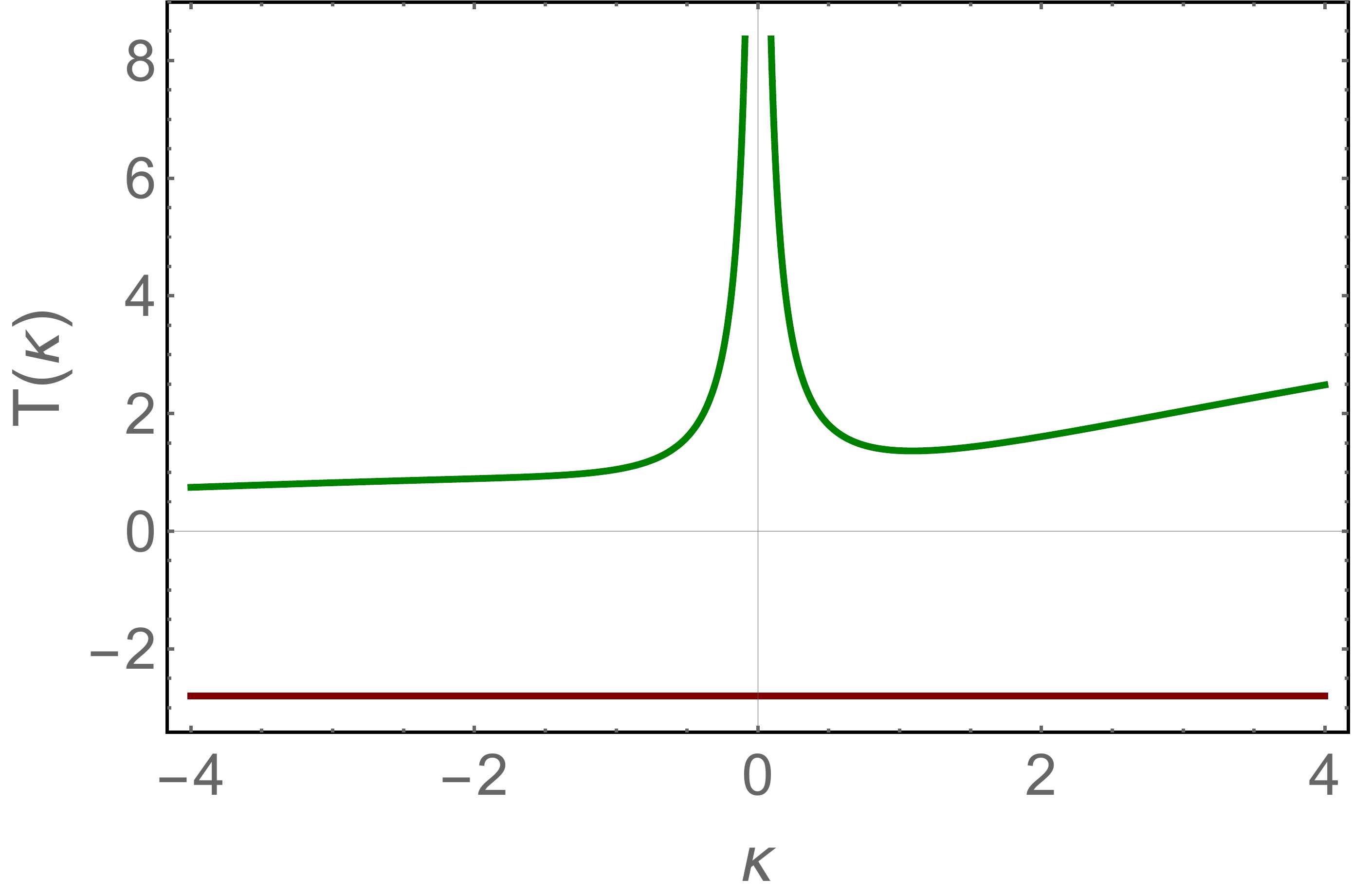} &~~~~&      \includegraphics[width=8.5cm]{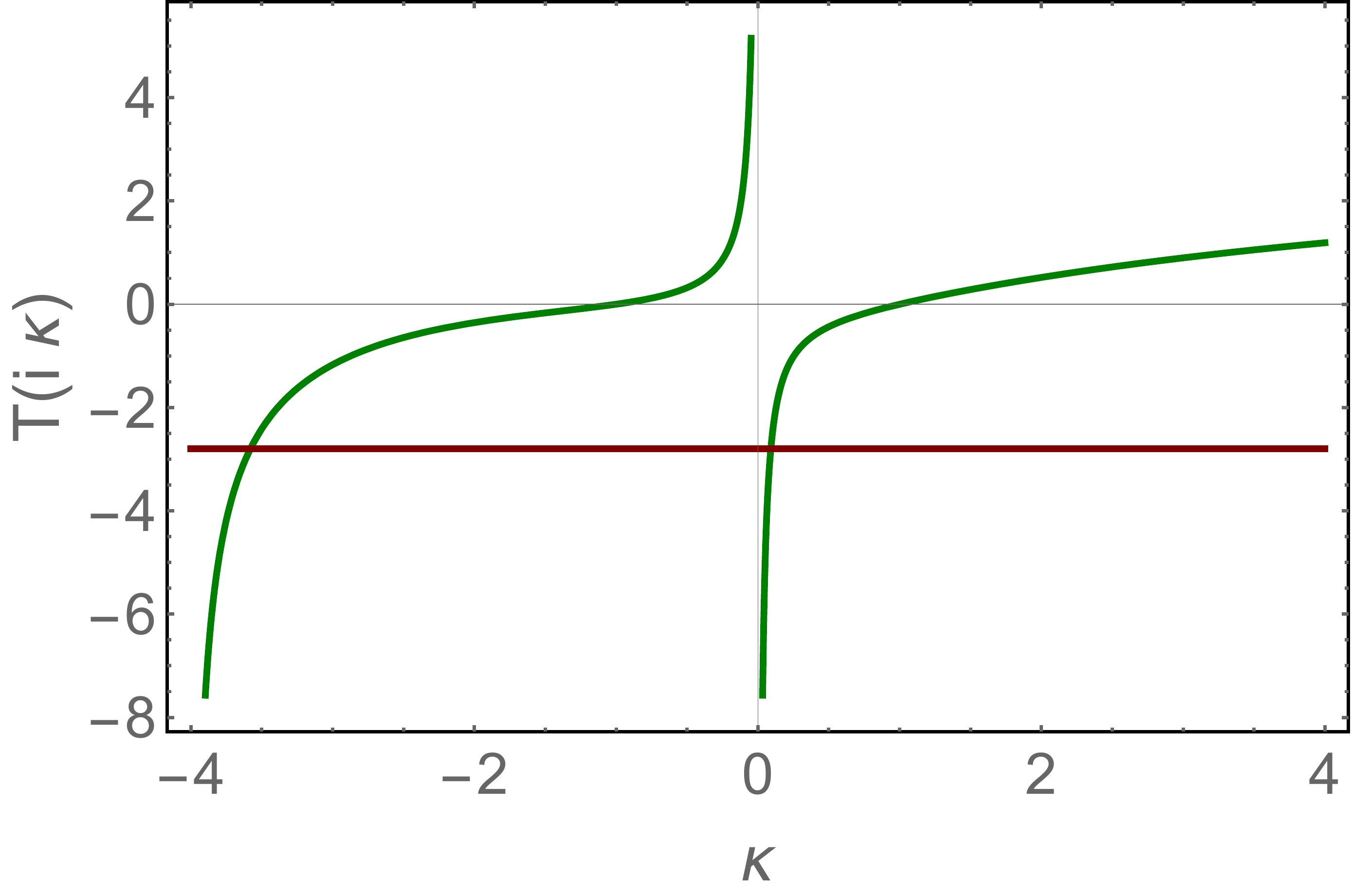}\\
     \fig{sigma1}-a & &\fig{sigma1}-b\\
     \end{tabular}      
     \end{center}    
      \caption{$\omega_G\Lb \kappa\Rb$ versus $\kappa$ for the Gribov 
propagator
 of \eq{GLPR}. The red line corresponds to $T\Lb \kappa\Rb = - \Delta_{\rm
 BFKL}\,\,=\,\,- 4 \ln 2$, where $\Delta_{\rm BFKL}$ is the intercept of
 the BFKL Pomeron for QCD. \fig{sigma1}-a: $T\Lb \kappa\Rb$, 
\fig{sigma1}-b:$T\Lb i\,\kappa\Rb$. }
      \label{sigma1}
   \end{figure}

    At first sight, the behaviour of the kinetic energy (see \fig{sigma} )
 for the BFKL  with Gribov's confinement, is not  that  different from the
 case that we have described in section III.  Indeed,  $T $ is negative 
 for negative $\kappa$ ( see \fig{sigma}-a), and due to this, we expect
  we have a bound state as in the case of the model, discussed
 in section III. As in the model  of section III  for negative  $\kappa$,
 we expect the  eigenfunction, which is small at large $r$ 
($\Psi\,\propto\,
\exp\Lb - \sqrt{|\kappa|}\,r\Rb$). Hence, we expect that the scattering
 amplitude   will decrease at large $b$. For example, we see  such a 
situation
  in \fig{sigma}-b, where the kinetic energy is plotted for the gluon
 propagator, which is in agreement with lattice QCD data \cite{DOS}.
 However, the actual setup is more  interesting: for the 
propagator
 of \eq{GLPR}, the kinetic energy is positive for all values of
 $\kappa$ (see \fig{sigma1}). We will argue below that in this
 case, the generalized BFKL Pomeron has the intercept which is 
equal to zero, and the eigenfunctions do not decrease exponentially
 at large $r$.

     In the appendix A  we discuss the $\kappa$ dependence of $\omega_G$ in
 more detail.

   ~

     \begin{figure}[ht]
     \begin{center}
     \begin{tabular}{ccc}
     \includegraphics[width=5.8cm]{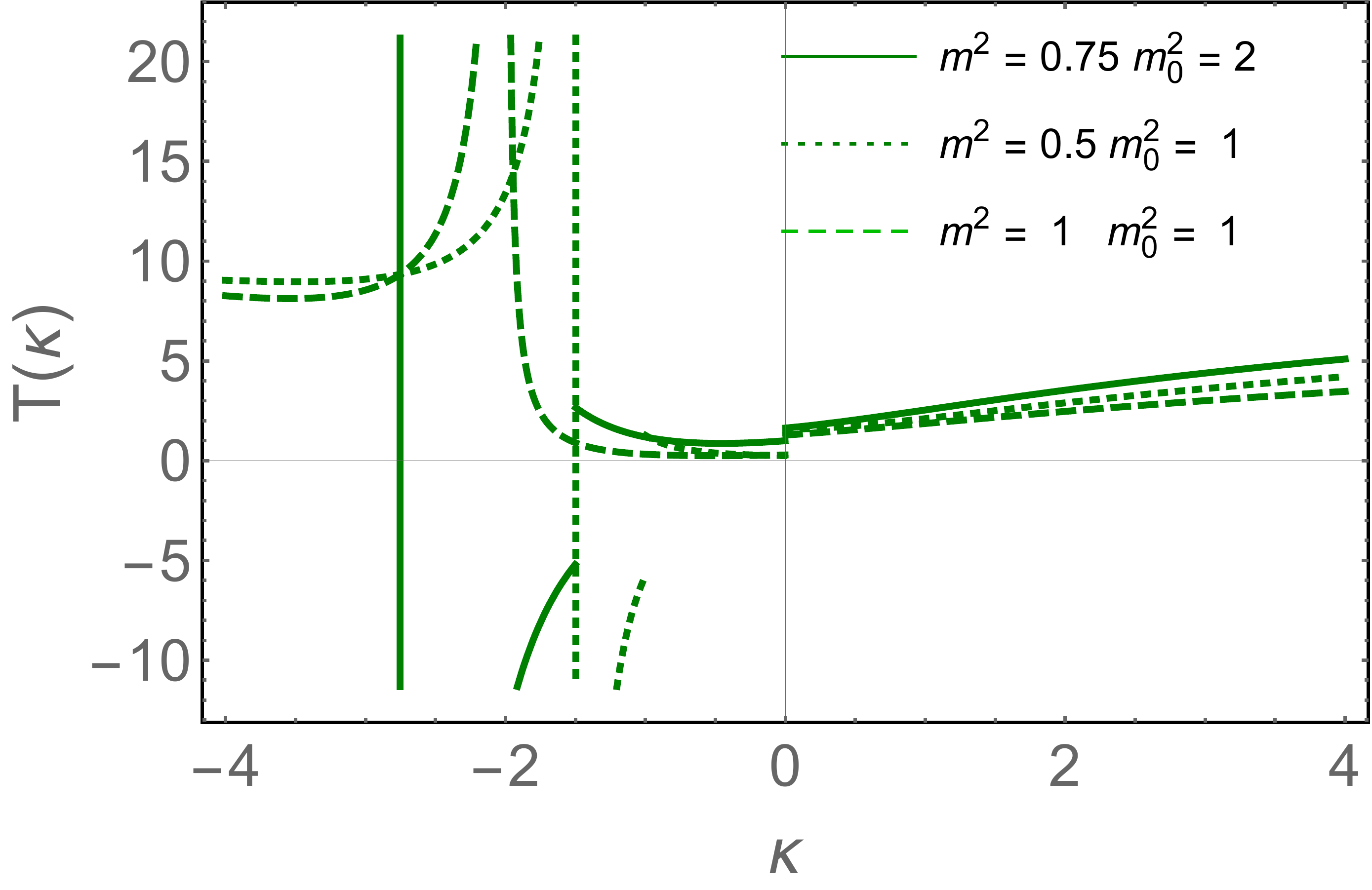} &    \includegraphics[width=6.cm]{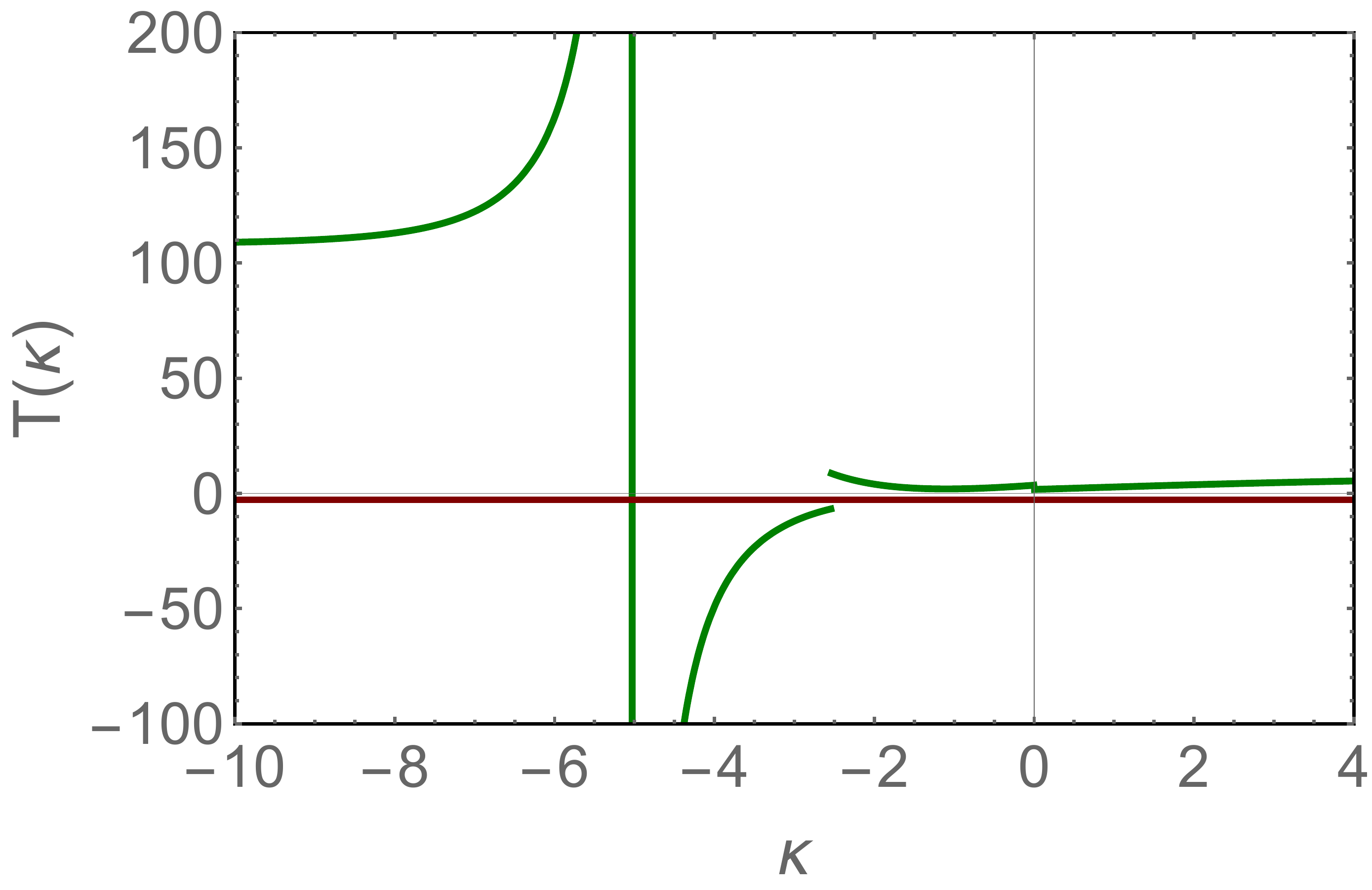}&  \includegraphics[width=5.8cm]{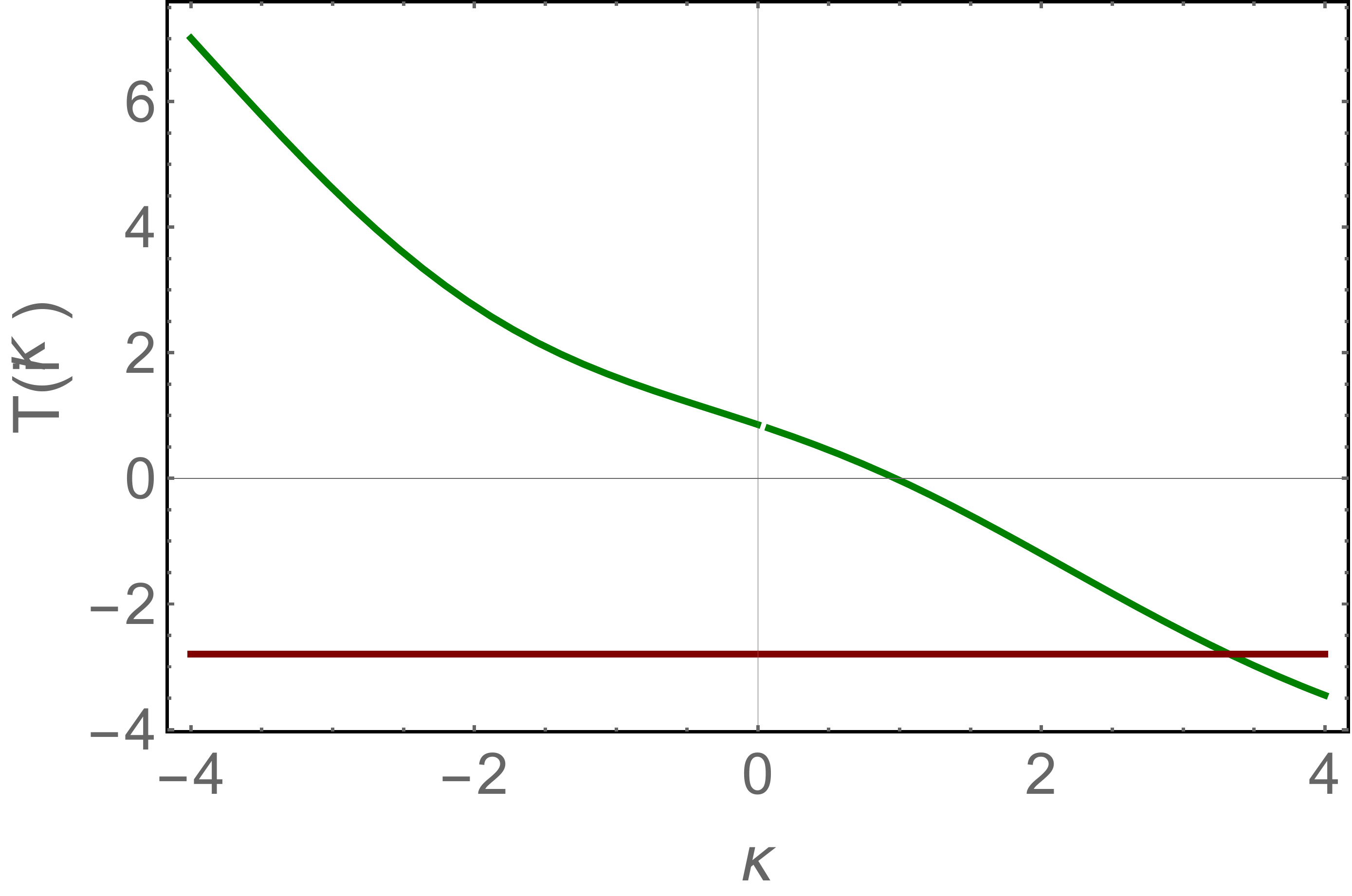} \\
     \fig{sigma}-a &\fig{sigma}-b&\fig{sigma}-c\\
     \end{tabular}
     \end{center}    
      \caption{\fig{sigma}-a: $T\Lb \kappa\Rb$ versus $\kappa$ at 
different values of $m$ and $m_0$.\fig{sigma}-b: The same as \fig{sigma}-a
 but for  $m_0=5.03$ , $m=  1.27$ and $\mu^2=0.459\,GeV^2$, which
 corresponds to the description of recent  calculations in lattice
 QCD(\cite{DOS}, section 2.1).  In \fig{sigma}-c the kinetic energy
 $T\Lb i \,\kappa\Rb$ is plotted. The red line corresponds to $T\Lb
 \kappa\Rb = - \Delta_{\rm BFKL}\,\,=\,\,- 4 \ln 2$, where
 $\Delta_{\rm BFKL}$ is the intercept of the BFKL Pomeron for QCD.}
\label{sigma}
   \end{figure}
  
   
     \begin{boldmath}
\subsection{ The BFKL  equation in momentum representation. }
\end{boldmath}

\begin{figure}[ht]
    \centering
  \leavevmode
      \includegraphics[width=16cm]{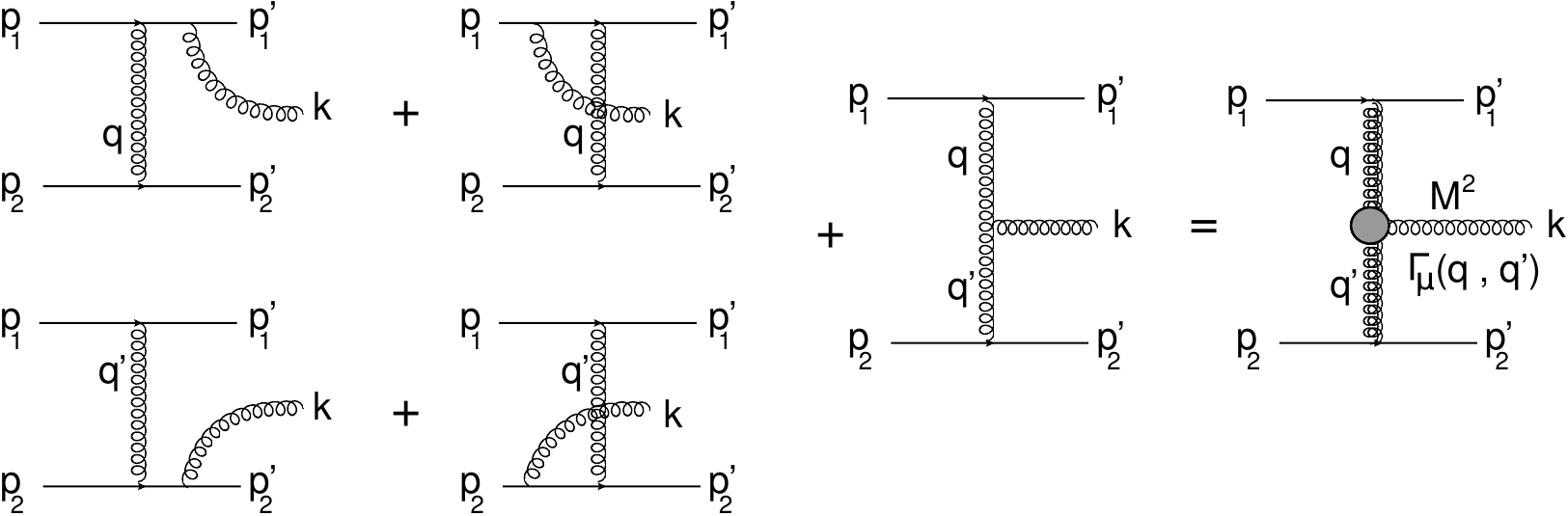}  
      \caption{The first Feynman diagrams with gluon emission, whose  sum
 leads to  $\Gamma_\nu\Lb q, q'\Rb$ (Lipatov vertex is denoted by the gray
 blob ). }
\label{3}
   \end{figure}

 In the previous section we  found $\omega_G\Lb q\Rb$, now we are 
going
 to find the kernel which is responsible for  gluon emission. Using
 the decomposition of \eq{GLPR} for the Gribov propagator, we can treat the production of the
 gluon as sum of two sets of the diagrams  (see \fig{3} )
 with $\tilde{M}^2 = \,\,i\mu^2$ and with $\tilde{M}^2  = -\,\,i\mu^2$.

We sum the first diagrams of the gluon emission shown in 
 \fig{3} to find the vertex $ \Gamma_\mu\Lb q, q'\Rb$ for
 the kernel of the BFKL equation (see \fig{eq}-b). It is
 easy to see that the sum shown in \fig{3}, leads to the
 Lipatov vertex that has the following form

\beq \label{V}
 \Gamma_\mu\Lb q, q'\Rb\,\,=\,\,-  q^\perp_{\mu}\,-\,q'^\perp_{\mu} \,+\,p_{1, \mu}\Lb - G^{-1}\Lb q\Rb\frac{1}{p_1 \cdot k}\,+\,\frac{p_2\cdot k}{p_1\cdot p_2}\Rb\,-\,p_{2, \mu}\Lb -   G^{-1}\Lb q'\Rb\frac{1}{p_2 \cdot k}\,+\,\frac{p_1\cdot k}{p_1\cdot p_2}\Rb 
 \eeq

 The gluon production vertex for the conjugated reggeized gluon can be
 written as follows
 \beq \label{VC}
 \tilde{\Gamma}_\mu\Lb q_1, q_1'\Rb\,\,=\,\, -
  q^\perp_{1,\mu}\,-\,q'^\perp_{1,\mu} \,+
\,p_{1, \mu}\Lb - G^{-1}\Lb q_1\Rb\frac{1}
{p_1 \cdot k}\,+\,\frac{p_2\cdot k}{p_1\cdot p_2}\Rb\,-\,p_{2, \mu}\Lb -
   G^{-1}\Lb q'_1\Rb\frac{1}{p_2 \cdot k}\,+\,\frac{p_1\cdot k}{p_1\cdot p_2}
\Rb \eeq

 The BFKL kernel for one   given configuration of the
 masses ( say, $\tilde{M^2}\,=\,+ i \,\mu^2$) at 
 $Q_T = 0$\footnote{$Q_T$ is the momentum transferred by the BFKL
 Pomeron, a conjugate variable to the impact parameter.} ( for
 forward scattering $q = q_1, q' = q'_1$) is given by
  \beq \label{KER} 
 \Lb G^{-1}\Lb q'\Rb\Rb^2 K\Lb q, q'\Rb\,=\, 
  \Gamma_\mu\Lb q, q'\Rb \cdot \tilde{\Gamma}_\mu\Lb q, q'\Rb\,=\\
\,\,\underbrace{4\Bigg(\frac{G^{-1}\Lb q\Rb \,G^{-1}\Lb q'\Rb}{k^2
 + \tilde{M}^2}\Bigg)}_{\mbox{gluon emission}}\, \, -\,\,\underbrace{2\,M^2\Lb 
\frac{N^2+1}{N^2_c}\Rb}_{\mbox{contact term}}\eeq 
    where the contact term has been discussed in Ref.\cite{LLS} and
 $K\Lb q, q'\Rb$  is the BFKL kernel  of gluon emission.  $N_c$ in
 \eq{KER} denotes the number of colours.
    
   Illustrating the derivation of \eq{KER}, we  calculate the diagram
 with the emission of one gluon in quark-antiquark scattering, to 
understand
 the structure of the BFKL equation (see \fig{3}). The contribution of
 this diagram is equal to
\beq\label{GE1}
N\Lb\fig{3}\Rb\,=\,\frac{\as N_c}{2 \pi^2}\int d^2 q \,d^2 q' \,G^2\Lb q\Rb \,   \Gamma_\mu\Lb q, q'\Rb \cdot \tilde{\Gamma}_\mu\Lb q, q'\Rb\,G^2\Lb q'\Rb \nn
\eeq

The gluon emission term can be re-written in the simple form
\beq \label{GE2}
N_{\mbox{gluon emission}}\Lb\fig{3}\Rb\,\,=\,\,\frac{\as N_c}{2 \pi^2}\int
 d^2 q \,d^2 q'\,G\Lb q\Rb\,G\Lb k\Rb G\Lb q'\Rb\eeq

Collecting all terms, including the gluon reggeization, which has
 been discussed in the previous section, and
  summing the contributions with $M^2$ and $\tilde{M}^2$,
 we obtain the BFKL equation
 in the form:
\beq \label{BFKLMR}
\omega \,\phi\Lb \omega, q\Rb\,\,=\,\,-\, 2 \omega_G\Lb q\Rb\,\,\phi\Lb 
\omega, q\Rb\,\,+\,\,\bas \int \frac{d^2 q'}{\pi}  G\Lb \vec{q} -
 \vec{q}^{\,'}\Rb \,\,\phi\Lb \omega, q'\Rb\eeq

Assuming that $\phi\Lb q\Rb$   depends only on $|\vec{q}|$, we can
 integrate the emission kernel over the angle and 
  in terms of  the variable of \eq{VAR1}, \eq{BFKLMR} takes the 
form:
\beq 
\label{BFKLMR1}
E \,\phi\Lb \kappa\Rb\,\,=\,\,\,  T\Lb \kappa\Rb\,\,-\,\, \int d \kappa' \,K\Lb \kappa ,\kappa'\Rb\,\phi\Lb \kappa'\Rb \,\,\,
\eeq

where 

\beq  \label{BFKLMR10}
K\Lb \kappa,  \kappa'\Rb\,\,\,=\,\,\,\displaystyle{{\rm Re}\Bigg\{\frac{1 \,+\,i\,m_0}{\sqrt{2 (m+i) (\kappa\,\,+\,\,\kappa')+(m+i)^2+(\kappa\,-\,\kappa')^2}}\Bigg\} }
 \eeq
 and 
\beq \label{BFKLMR2}
G\Lb \kappa\Rb\,\,=\,\,\frac{ \kappa \,+\,m\,+\,m_0}{\Lb \kappa\,+\,m\Rb^2\,\,+\,\,1};~~~~ \kappa \,=\,q^2/\mu^2
\eeq
 In \eq{BFKLMR1}    - \eq{BFKLMR2} we introduce $m$ and $m_0$,
 which are equal to $m^2/\mu^2$ and $m^2_0/\mu^2$, respectively
 (see \eq{VAR1}).

This equation appears to be  similar to the BFKL equation  for a massive 
gluon 
(see Ref. \cite{LLS}, and section III)  in  the non-abelian 
Yang-Mills theories with a Higgs particle, which is responsible for
   generation of mass.
 However, we do not have a contact term in \eq{BFKLMR},  which stems
 in such an approach from the mass of the gluon and from  Higgs
 production. It is instructive to note, that for the Gribov's propagator
 the contact term does not  appear, even if we assume the existence of a
 Higgs meson, with  mass squared $\pm i \mu^2$. A more general form of 
the
 gluon propagator, which is given in \eq{GGLPR1} and which we view as
 a sum of Gribov's propagators, also does not generate a contact term.
  Therefore, the absence of  a contact term in our equation, is a direct 
 indication that Gribov-Zwanziger confinement does not lead to  a massive
 gluon.

 ~


\subsection{The Pomeron intercept. }
\subsubsection{General features of the equation's spectrum}
Following the general pattern of Ref.\cite{LLS} we can re-write 
 \eq{BFKLMR1}   in the form of \eq{DIS5} -\eq{P} (see section IIIE.2):
\beq \label{HGR}
E\,\Psi\Lb r \Rb\,\,=\,\,{\cal H}\,\Psi\Lb r \Rb
\eeq
with
\beq\label{HGR1}
{\cal H}\,\,=\,\,T\Lb \hat{\kappa} \Rb \,\,-\,\,G\Lb r \Rb \,\,+\,\,\frac{N^2_c + 1}{2\,N^2_c} \hat{P}\eeq
where  $\sqrt{\hat{\kappa}} \,\,=\,\,-\,i \vec{\nabla}_\perp$ and
  $\hat{P}$ is  a shorthand notation for the projector onto the
 state $\sim m^2/(p^2+m^2)$
\beq \label{PGR}
\hat{P} \,\phi\Lb p \Rb\,=\,2 m^2 G\Lb \kappa\Rb \int d \kappa'
  G\Lb \kappa'\Rb \phi\Lb \kappa'\Rb\,\,\xrightarrow{\rm coordinate\,
\,image} G\Lb r \Rb \,C_1 ~~~\mbox{where} ~~C_1\,=\,2 m^2\int d \kappa'
  G\Lb \kappa'\Rb \phi\Lb \kappa'\Rb
 \eeq
For large $r$,  $G\Lb r\Rb$ exponentially decreases (see \eq{GLPR3}) as
 well as the contact term. Hence, at large $r$ \eq{HGR} takes the
 following form:

\beq \label{HGR2}
E\Psi\Lb r \Rb \,\,=\,\,T\Lb \hat{\kappa} \Rb \Psi\Lb r \Rb
\eeq
with the eigenfunctions of \eq{PLWA}. Denoting the large asymptotic behaviour
 of the eigenfunction as $\Psi\Lb r \Rb \xrightarrow{r\,\gg\,\,1/\mu} \,
\,\exp\Lb - \sqrt{a}\,r\Rb$, we see that the energy is equal to
\beq \label{HGR3}
E \,\,=\,\,T\Lb-\, a\Rb
\eeq
On the other hand,    it is shown in Ref.\cite{LLS}
 (see section III-D\footnote{In Ref.\cite{LLS}  it is demonstrated
  that for a rather general form of the wave function,  the typical
 $\kappa'$ in the integral $\int d\, \kappa'\,K( \kappa, \kappa')
 \,\phi(\kappa')$, is $\kappa'  \approx \kappa$ for $\kappa\,\,
 \gg\,\,1$. Note, that in Eq. 51 of  Ref.\cite{LLS}  there is
 a missprint: $\beta$ should be replaced by $x$ in the denominator.}) 
 that in the region of small $r$  \eq{HGR1} reduces to 
 the massless QCD BFKL equation \cite{FKL,BFKL,LIP}:

\beq \label{HGR4}
E\,\Psi\Lb r \Rb \,\,=\,\,{\cal H}_0 \,\Psi\Lb r \Rb
\eeq
where\cite{LIP}
\beq \label{H0}
{\cal H}_0\,=\,\ln p^2 \,+\,\ln |r|^2\,-\,2 \psi\Lb 1 \Rb\,
\eeq
The  eigenfunctions of \eq{HGR4} are $\Psi\Lb r \Rb \,=\,r^{2(1- \gamma)}
 $, and   the  eigenvalues of \eq{HGR4}   can be parametrized as a 
function 
of $\gamma$ (see \eq{CHI}). Therefore, for $r\,\to\,0$ we have the
 eigenvalue which is equal to
\beq \label{HGR5}
E\, \,\,=\,\,\chi\Lb \gamma\Rb
\eeq
From \eq{HGR3} and \eq{HGR5} we can conclude, that the value of $a$ 
and $\gamma$ are  correlated, since
\beq \label{HGR6}
E\, \,\,=\,\,\chi\Lb \gamma\Rb\,\,=\,\,T\Lb -\, a \Rb
\eeq
Based on \eq{HGR6} we expect that the minimum eigenvalue is equal
 to $\chi(\h) \,=\,-\,4\,\ln 2$. For the simplest Gribov's propagator
 of \eq{GLPR} we see from \fig{sigma1} that  $T\Lb \kappa\Rb \,>\,0$ 
 for all values of $- \infty \,<\,\kappa\,<\,+\infty$ which
 means that $\gamma$ in \eq{HGR6}, should be such that $\chi\Lb 
\gamma\Rb>0$.
 Consequently, we  infer that \eq{HGR6} contradicts \eq{HGR5}.

     \begin{figure}[ht]
     \begin{center}
     \includegraphics[width=9.5cm]{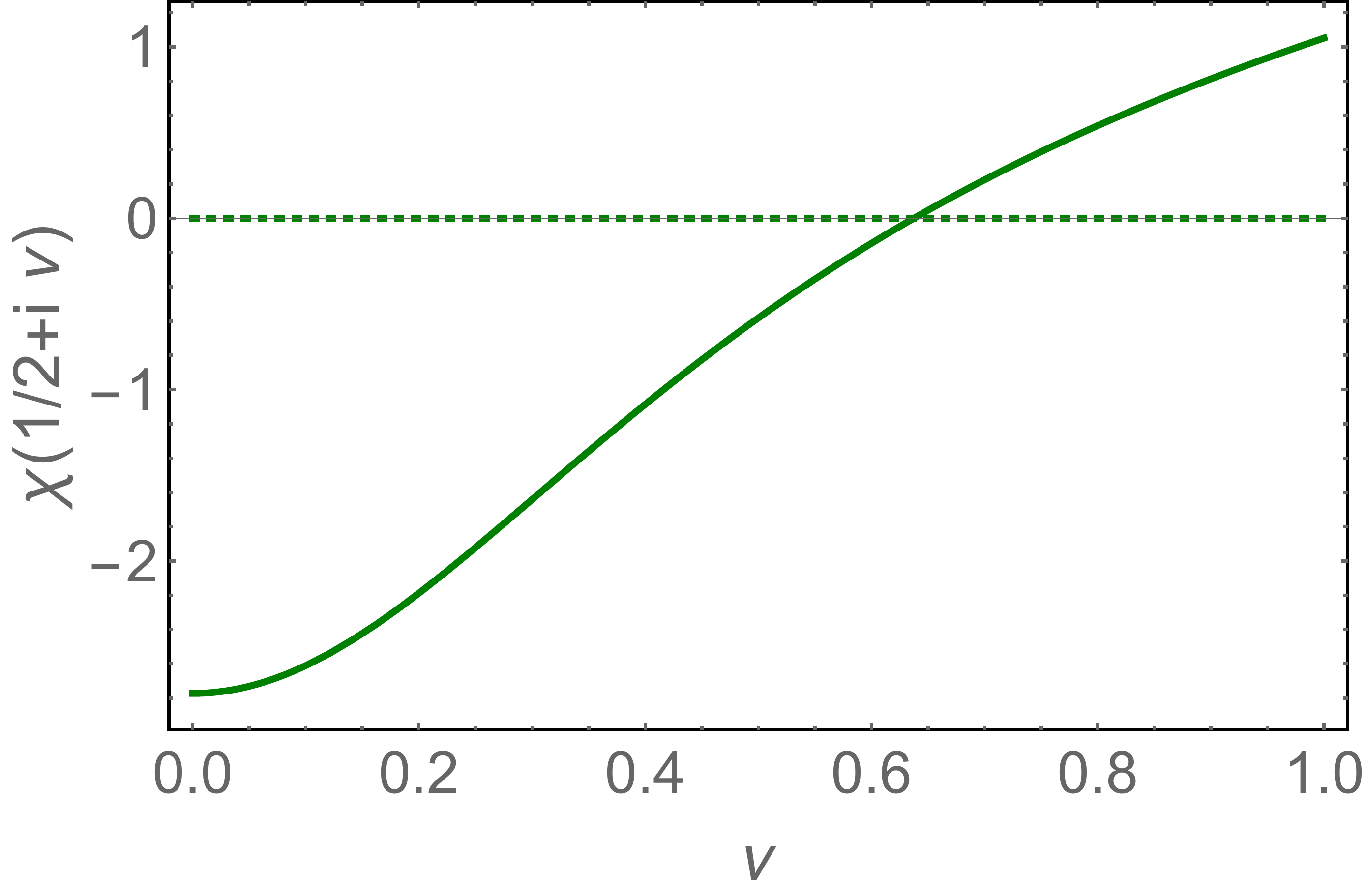}   
     \end{center}    
      \caption{The eigenvalues of the BFKL equation for QCD 
 $E\,=\,\chi( \gamma)$ versus $\gamma= \h  \,+\,i\,\nu$. The
 solid line shows the real part of E, while the dotted line shows the 
imaginary part.}
      \label{chi}
   \end{figure}   

    A possible  way out of this contradiction, could be that both
 equations are correct for specific values of $\gamma$.
     In \fig{chi} we plot the eigenvalues of the massless BFKL equation
 for $\gamma = \h \,+\,i\,\nu$.  One can see that for $\nu \,\geq\,0.6$  
the
 energy is positive.  
 Hence, this value of $\nu$ could correspond to the
 Pomeron with the intercept which is equal to zero.  The fact that the
 so called soft Pomeron has a small intercept, is one of the reliable
   results of the high energy phenomenological attempt to describe the 
soft
 data at the LHC.  
   
   However, for $m \,\neq\,0$ and $m_0 \,\neq\,0$  the kinetic energy
 could be negative, and  \eq{HGR6} holds for $\gamma = \h$,  leading to
 the intercept of the Pomeron which coincides with the intercept of the
 massless BFKL Pomeron. In particular, this is the case for the gluon
 propagator which describes the lattice QCD data ( see Ref.\cite{DOS}
 and \fig{sigma}-b).

 \subsubsection{Estimates from the variational method}
  As we have discussed above, we expect that  (1) the energy of the
 ground state will be close to zero for  $m \,=\,0$ and $m_0 \,=\,0$;
 and (2) it
   will be the same as for the massless BFKL equation, for  $m^2 
\,\neq\,0$
 and $m^2_0 \,\neq\,0$ .   In this section we check this, using 
 the variational  approach. 
 In this approach, the upper bound for the ground state energy $E_{0}$ of 
the Hamiltonian $\mathcal{H}$ may be found  by minimizing the functional
 \beq \label{VP1}
 E_{\mbox{ground}}\,\,\equiv\,\,E_0\,\,\,\leq \,\,\,
 F[\{\phi\}]\,\,=\,\,\frac{\Big{\langle} \phi^{\ast}\!\Lb r \Rb \Big{|} {\cal H}\Big{|} \phi\!\Lb r\Rb \Big{\rangle}}{\Big{\langle} \phi^{\ast}\!\Lb r \Rb \Big{|} \phi\!\Lb r\Rb \Big{\rangle} }
 \eeq
  \eq{VP1} means that the functional $F[\{\phi\}]$ has a minimum for the
 function $\phi_0\Lb r \Rb$, which is the eigenfunction of the ground
 state with energy $E_0$.

 For our Hamiltonian in  momentum space, \eq{VP1} can be re-written in
 the form
\bea\label{VP2}
&& E_{0}\,\,=\\
&&\min_{\phi}\left\{\frac{\int^\infty_{0} d \kappa \, T\Lb \kappa\Rb \, | \phi\Lb \kappa\Rb|^2  \,-\,\int^\infty_0 d \kappa \int^\infty_0  d \kappa' \,K\Lb \kappa,\kappa'\Rb\,\phi\Lb \kappa\Rb \phi^*\Lb \kappa'\Rb \,
  \,+\,2 m\frac{N^2_c + 1}{2 N^2_c} \Big{|} \int^\infty_0  d \kappa\, \frac{\phi\Lb \kappa\Rb }{\kappa + 1}\Big{|}^2}{\int^\infty_0 d \kappa  | \phi\Lb \kappa\Rb|^2}\right\}\nn
\eea

 The success of finding the value of $E_0$, depends on the choice of the
 trial functions in \eq{VP2}.  We choose it in the form
 \beq \label{TRIF}
 \phi_{\mbox{trial}}\Lb \kappa\Rb\,\,=\,\,\Lb \frac{\kappa}{ \kappa^2 + a^2}\Rb^{\gamma}
 \eeq 
 
 In the coordinate representation~\eq{TRIF} corresponds to
 \bea \label{TRIFCR}
\Psi_{\mbox{trial}}\Lb r \Rb\,\,&=&\,\, 2^{-2 (\gamma +2)} \left(\frac{1}{a^4}\right)^{-\gamma } \left(a^4\right)^{-\gamma } \left(\frac{\sqrt{\pi } 2^{\gamma +3} \left(\frac{1}{a^4}\right)^{\frac{\gamma -1}{2}} \Gamma \left(\frac{\gamma -1}{2}\right) \, _1F_4\left(\frac{\gamma }{2}+\frac{1}{2};\frac{1}{2},\frac{1}{2},1,\frac{3}{2}-\frac{\gamma }{2};-\frac{1}{256} a^4 r^4\right)}{\Gamma \left(\frac{\gamma }{2}\right)}\right.\nn\\
 &  &\left.+\frac{32 \Gamma (1-\gamma ) r^{2 \gamma -2} \, _1F_4\left(\gamma ;\frac{\gamma }{2}+\frac{1}{2},\frac{\gamma }{2}+\frac{1}{2},\frac{\gamma }{2},\frac{\gamma }{2};-\frac{1}{256} a^4 r^4\right)}{\Gamma (\gamma )
 }\right.\nn\\
 &&\left.
 -\frac{\sqrt{\pi } 2^{\gamma } \gamma  r^2 \left(\frac{1}{a^4}\right)^{\frac{\gamma }{2}-1} \Gamma \left(\frac{\gamma }{2}-1\right) \, _1F_4\left(\frac{\gamma }{2}+1;1,\frac{3}{2},\frac{3}{2},2-\frac{\gamma }{2};-\frac{1}{256} a^4 r^4\right)}{\Gamma \left(\frac{\gamma +1}{2}\right)}\right) \\
 &&
 \rightarrow\,\,\left\{ \begin{array}{l l} r \,\rightarrow \infty\,\, & \,\,\, \propto ~~r^{-2 (\gamma + 1)} \\
  &\\
  r \,\rightarrow\,\, 0 &\,\,\, \propto\,~~r^{-2 + 2 \gamma} \end{array}\right.
  \eea
  
   The form of the trial function  was suggested  by the form of 
Gribov's
 propagator.
 One can see that our trial function has the  expected behaviour
 for the case of $m=0, m_0 =0$ , if $a >0$
 and $ b\,=\,2 \ga - 1\,>\,0$, leading  to a power-like decrease at large  
$r$.
  Such a function cannot  be an eigenfunction of ${\cal H} = T(\kappa)$, 
indicating possible difficulties with \eq{HGR6}.  
 
\begin{figure}[ht]
     \begin{center}
     \begin{tabular}{ccc}
     \includegraphics[width= 6cm]{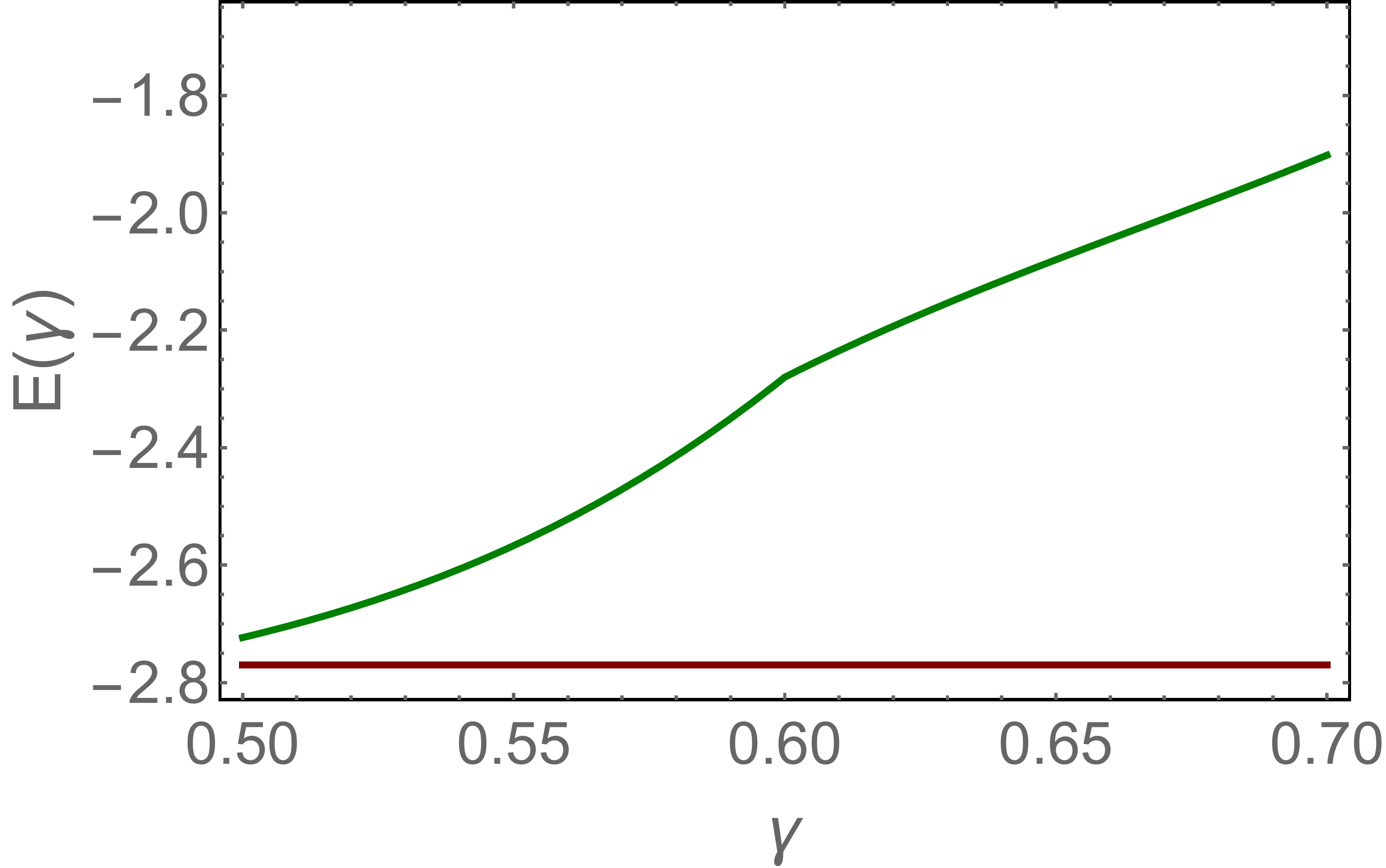} &    \includegraphics[width=5.9cm]{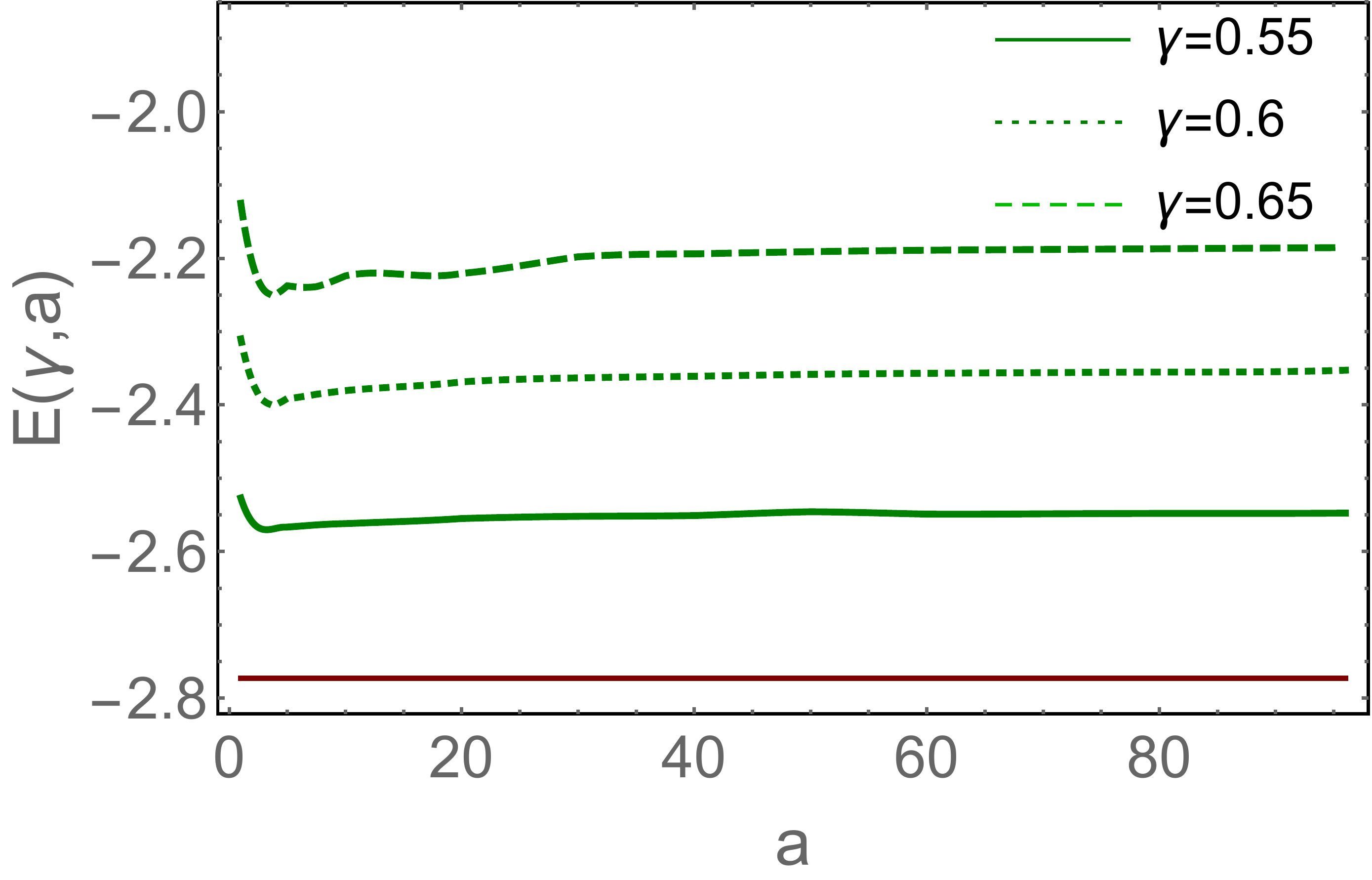}& \includegraphics[width=6.cm]{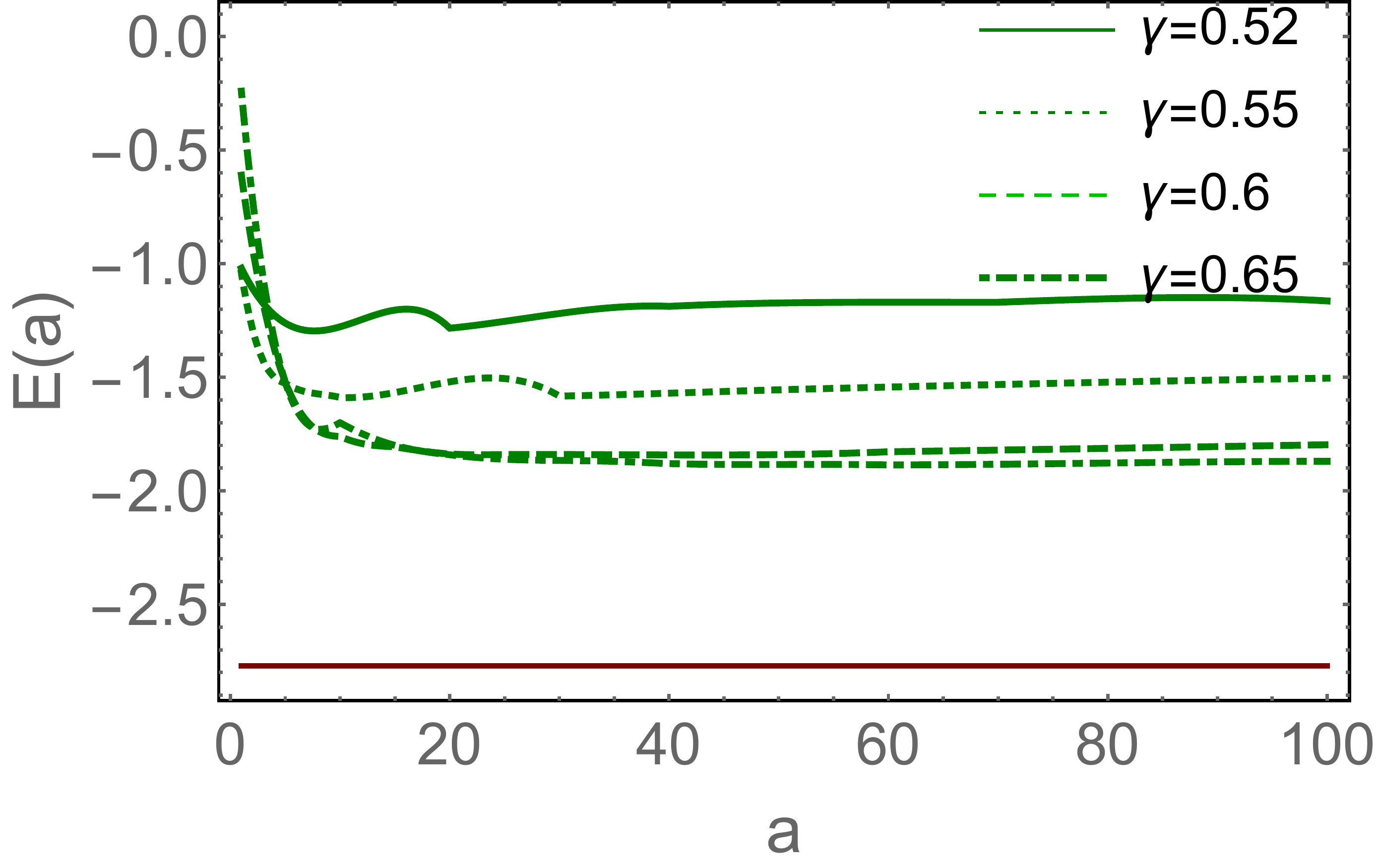}\\  
       \fig{egr}-a & \fig{egr}-b  &\fig{egr}-c\\
     \end{tabular}
     \end{center}    
      \caption{\fig{egr}-a:  $E \equiv  - {\rm Pomeron \,\,intercept}$ of
 the BFKL Pomeron  versus $\gamma$ at different values of $\gamma$ for 
 $a = 100$ and $m =0$ and $m_0=0$. These values of $m$ and $m_0$
 correspond to the  Gribov's propagator of \eq{GLPR}, which stems
 from Refs.\cite{GRI0,GRI1,GRI2,GRI3,GRI4}. \fig{egr}-b: The same
 as in \fig{egr}-a but at fixed values of $\gamma$ as function of
 $a$. In  \fig{egr}-c we plot  the energy as function of $a$  for
 the case of  $m =1.27\mu^2 $ and $m_0=3.76 \mu^2$ , which describes
 the lattice QCD evaluation of the gluon propagator \cite{DOS}. The
 estimates for $E$ were performed in the framework of the  variational
 method with the trial function $\phi_{\rm trial}$ of \eq{TRIF} (see
 appendix B for details).}
\label{egr}
   \end{figure}

 In \fig{egr}-a and \eq{egr}-b  we calculate $ 
 E_{\mbox{ground}}$ from \eq{VP1}, for the case of Gribov's propagator of
 \eq{GLPR} ($m=0,m_0=0$). In appendix B we describe the details of the
 numerical estimates.

 We obtain  the minimal energy corresponds
 to $\gamma\,=\,\h$ in accord with our expectation. However,
  $E_{\mbox{ground}}\,\,\to\,\,- 4\ln 2$  instead of $E_{\mbox{ground}}
 = 0$, in contradiction to our expectations.  Note that the
 singularities of the trial function corresponds to $\kappa \,=\,i\,a$.
 In \fig{sigma1}-c we plot the kinetic energy at pure imaginary $\kappa$,
 and we  see that $T\Lb i \kappa \Rb$ can be negative and  
 equal to $- 4\,\ln 2$. 
 
 For   $m^2 \,\neq\,0$ and $m^2_0 \,\neq\,0$  we face a different
 problem: $\gamma$ turns out  to be larger than expected $\gamma=\h$,
 and the energy level is far away from the ground state energy
 $E = - 4 \ln 2$   for the massless BFKL equation(see \fig{egr}-c).
 Perhaps, this result is due to our choice of the trial function, not
 being satisfactory. We believe, that both observations show that we need 
to
 solve \eq{BFKLMR1} numerically, in the same way as it has been
 done in Ref.\cite{LLS}. We intend to do this in the near
 future, and we will publish the results elsewhere.
  
\begin{boldmath}
\subsection{ The BFKL  kernel in the coordinate representation. }
\end{boldmath}

   Using \eq{CORE1}, and the decomposition of \eq{GGLPR3}, we  obtain
 the gluon propagator of \eq{GGLPR1} in the coordinate representation
 in the form:
 \beq   \label{GLCO}
 G\Lb r\Rb\,\,\,=\,\,\,  \frac{1}{2} \left(\left(1-\frac{\text{m}_0^2}{i}\right) K_0\left(r\,\sqrt{m^2-i}\right)+\left(1+\frac{\text{m}^2_0}{i}\right) K_0\left(r\,\sqrt{m^2+i} \right)\right)   
  \eeq 
  
   Note, that we now return to using the notation $m = M/\mu$ and $m_0 = 
\sqrt{M^2_0\,-\,M^2}/\mu$ .
  
   At large $r$ it tends to
   \bea   \label{GLCO1}   
 G\Lb r \Rb&\,\,\xrightarrow{\mu\,r\,\gg\,\,1} & \h \sqrt{\frac{\pi }{2\,\,r}} {\rm Re}\left(\frac{\left(1-i \text{m}^2_0\right) e^{-\,r\,\sqrt{m^2+i} }}{\sqrt[4]{m^2+i}}\right) \\
 &=&  \h \sqrt{\frac{\pi }{2\,\,r}}   e^{ - r  \sqrt{m^2\,+\,1} \cos \theta}\Bigg( \cos\Lb - r  \sqrt{m^2\,+\,1} \sin \theta  + \h\theta\Rb
 - \,m^2_0 \sin\Lb - r  \sqrt{m^2\,+\,1} \sin \theta  + \h\theta\Rb\Bigg)\nn
 \eea
 where
  $$\cos\theta \,\,=\,\,\sqrt{ \h \Lb 1\,\,+\,\, \frac{m}{\sqrt{m^2\,+\,1}}
 \Rb}; ~~~\sin\theta \,\,=\,\,\sqrt{ \h \Lb 1\,\,-\,\, \frac{m}{\sqrt{m^2\,
+\,1}} \Rb} .$$
 
   Hence, one can see that the gluon propagator decreases exponentially
 at large $r$, even at $m=0$.    From \eq{GLCO1} we can conclude that
   \beq \label{GLCO2}
   \Sigma\Lb r \Rb\,\,\,=\,\, G^2\Lb r \Rb   \xrightarrow{\mu\,r\,\gg\,\,1}  \Bigg( \h \sqrt{\frac{\pi }{2\,\,r}} {\rm Re}\left(\frac{\left(1-i \text{m}^2_0\right) e^{-\sqrt{m^2+i} r}}{\sqrt[4]{m^2+i}}\right) \Bigg)^2 \,\,\propto\,\,e^{ - \,r\,2 \sqrt{m^2 
 \,+\,1}\cos\theta}
 \eeq  
  From \eq{GGLPR1} and $\vec{q}\,=\,- i\,\vec{\nabla}$ we conclude that
 $\omega_G\Lb r\Rb$ is equal to
  \beq \label{GLCO3}
   \omega_G\Lb r\Rb\,\,=\,\,\int d^2 \,r'\,K_0\Lb M_0 \,| \vec{r}\,-\,\vec{r}'|\Rb \Big( \Lb -\nabla^2_{r'} \,\,+\,\,m^2\Rb^2 + 1\Big)\,\Sigma\Lb r'\Rb
   \eeq
    In \eq{GLCO3} the behaviour of   $\omega_G\Lb r\Rb$ at large $r$ stems
 from the integration in two regions: $
M_0 \,| \vec{r}\,-\,\vec{r}'|\,\leq\,1$ and $ \,r'\,2 \sqrt{m^2 
 \,+\,1}\cos\theta    \,\leq\,1$.  The first region leads to the asymptotic
  behaviour of \eq{GLCO3}, while the second region gives  
 $   \omega_G\Lb r\Rb \,\,\propto\,\exp\Lb - M_0\,r\Rb$.
 Hence for $M_0 \,<\,2 \sqrt{m^2 
 \,+\,1}\cos\theta$   $ \omega_G\Lb r\Rb \,\,\propto
\,\exp\Lb - M_0 \,\,r\,\Rb$. 
   Lattice QCD leads to  such  behaviour of $\omega_g$,  as
 it can be seen from \fig{sigma1}.      
       It should be stressed, that the  exponential decrease  depends on 
the
 value of the gluon propagator at $q =0 $  $G\Lb q=0\Rb\,\,
\propto\,M_0^2/\mu^4$.  In other words, the original Gribov
 propagator of \eq{GLPR} does not give the BFKL kernel which
 decreases exponentially at large $b$. Indeed, at $M_0 \,\to\,0$
 instead of $\exp\Lb - M_0 \,r\Rb$ decrease, we have $\ln r$ behaviour
 from the region $ \sqrt{m^2 + 1}\,r'\,\leq\,1$. However, even for 
       $m_0\neq 0$ one can see from \eq{GLPR}, that $\omega_G$  is a
 decreasing function, with  oscillations. 
    These oscillations do not contradict the unitarity constraints, 
  they also do  not violate the exponential decrease of the
 scattering amplitude at large $b$.

          ~

        ~


\section{Non-linear equation and the size of  Froissart disc. }


 The eigenfunctions  of the master  equation (see \eq{BFKLMR1})  at
 short distances are  proportional to $\Lb r^2\Rb^{1 - \gamma}$  and,
 therefore, for deep inelastic scattering, which occurs at short
 distances, the solution has the form of \eq{SFD2}. Hence, repeating
 the procedure that has been discussed in \eq{SP1}  -    \eq{GACR}, 
we obtain
 the same equations for the radius of the Froissart disc (see \eq{RFD}
 and \eq{RFD1}). The variable $\zeta $ takes the form: $\zeta =
 4 (m + m_0)b/\bas \,Y$. Actually, as we have discussed in section IV-E,
  the asymptotic  exponential decrease at $r \,\geq\,1/\mu$ is determined
 by the  smaller  of the two masses: $m + m_0$ and $2\,\sqrt{m^2 
\,+\,\mu^2}$. 
 For the realistic case of $m^2 = 1.27\mu^2$ and $m^2_0 = 3.76\,\mu$ 
\cite{DOS} $m + m_0\,\,<\,\,2\,\sqrt{m^2 \,+\,\mu^2}$.
 
 The non-linear equation has the same form as \eq{BK} with the kernel 
 \beq \label{KERFIN}
K\Lb \vec{r}', \vec{r} - \vec{r}'; \vec{r}\Rb \,\,\,=\,\,\frac{r^2}{r'^2}\int d^2 \,r''\,K_0\Lb M_0 \,| \vec{r}\,-\vec{r}' \,-\,\,\vec{r}''|\Rb \Big( \Lb -\nabla^2_{r''} \,\,+\,\,m^2\Rb^2 + 1\Big)\,\Sigma\Lb r''\Rb
   \eeq 
 It should be noted, that this kernel approaches the kernel of \eq{BK} at
 short distances.
 Generally speaking, at $ r \,\geq\,1/\mu$ (see \eq{GLPR}) we need to take
 into account the full kernel of \eq{KERFIN}. However, for DIS processes 
the typical $r \,\propto\,1/Q\,\,\ll\,\,1/\mu$, where $Q$ is the photon 
virtuality, and we can safely use the kernel of \eq{BK}, even in the 
saturation region, where $r^2\,Q^2_s\Lb Y,b \Rb\,>\,1$.  Restricting
 ourselves to the DIS process, we wish to consider the following $r$ 
 \beq \label{RINQ}
  \frac{1}{\mu^2}\,\,>\,\,r^2\,\,>\,\,\frac{1}{Q^2_s\Lb Y, b \Rb}
  \eeq
 
 However, even in this region the general non-linear 
  evolution of \eq{BK}
 is difficult to analyze analytically,  for the full BFKL kernel of
 \eq{KERC} . This kernel 
  includes the summation over all twist
 contributions. We would like to start  with a simplified version
 of the kernel in which  we restrict ourselves 
 to the leading twist term only\cite{LETU}. For the leading twist term 
we  only sum logs  terms, and actually we have two types of logs:
 $ \Big(\bas
 \ln\Lb r\,\Lambda_{QCD}\Rb\Big)^n$ in the perturbative QCD kinematic
 region where  $r\,Q_s\Lb Y,b\Rb\,\,\equiv\,\,\tau\,\,\ll\,\,1$; and
 $  \Big(\bas \ln\Lb r \,Q_s\Lb Y,b \Rb \Rb\Big)^n$ inside the
 saturation domain ($\tau\,\gg\,1$), where $Q_s\Lb Y, b \Rb$ denotes the
 saturation scale. To sum these logs  it is necessary to modify the 
BFKL kernel
 in different ways in the two kinematic regions, which takes the form 

\bea \label{SIMKER}
\chi\Lb \gamma\Rb\,\,=\,\, \left\{\begin{array}{l}\,\,\,\frac{1}{\gamma}\,\,\,\,\,\,\,\,\,\,\mbox{for}\,\,\,\tau\,=\,r Q_s\,<\,1\,\,\,\,\,\,\mbox{summing} \Lb \ln\Lb1/(r\,\Lambda_{\rm QCD})\Rb\Rb^n;\\ \\
\,\,\,\frac{1}{1\,-\,\gamma}\,\,\,\,\,\mbox{for}\,\,\,\tau\,=\,r Q_s\,>\,1\,\,\,\,\,\mbox{summing} \Lb \ln\Lb r Q_s\Rb\Rb^n;  \end{array}
\right.
\eea
instead of the full expression of \eq{KERC}. Recall that the solution
 has the form of \eq{SFD2}.

Inside  the saturation region where $\tau\,\,=\,\,r^2\,Q^2_s
\Lb Y,b\Rb\,\,>\,\,1$, the logs 
   originate from the decay of a large size dipole into one small
 size dipole  and one large size dipole.  However, the size of the
 small dipole is still larger than $1/Q_s$. This observation can be
 translated in the following form of the kernel
\beq \label{K2}
 \int \, \displaystyle{K\Lb \vec{x}_{01};\vec{x}_{02},\vec{x}_{12}\Rb}\,d^2 x_{02} \,\,\rightarrow\,\pi\, \int^{x^2_{01}}_{1/Q^2_s(Y,b)} \frac{ d x^2_{02}}{x_{02}^2}\,\,+\,\,
\pi\, \int^{x^2_{01}}_{1/Q^2_s(Y, b)} \frac{ d |\vec{x}_{01}  -
 \vec{x}_{02}|^2}{|\vec{x}_{01}  - \vec{x}_{02}|^2}
\eeq

Inside the saturation region \eq{BK} has the form
\beq \label{BK2}
\frac{\partial^2 \widetilde{N}\Lb Y; \xi, \vec{b}\Rb}
{ \partial Y\,\partial \xi}\,\,=\,\, \bas \,\left\{ \Lb 1 
\,\,-\,\frac{\partial \widetilde{N}\Lb Y; \vec{r}, \vec{b}
 \Rb}{\partial  \xi}\Rb \, \widetilde{N}\Lb Y; \xi,
 \vec{b}\Rb\right\}\eeq
where 
 $\widetilde{N}\Lb Y; \xi, \vec{b}\Rb\,\,=\,\,\int^{
r^2} d r'^2\,N\Lb Y; \vec{r}', \vec{b}\Rb/r'^2\,\,=\,\,\int^{\xi} d \xi' N\Lb Y,\xi',\vec{b}\Rb$ .
 
 Rewriting  $\frac{\partial}{\partial Y}   $ in terms of 
 \beq \label{Z}
z\,\,=\,\,- \,\frac{d\,\chi\Lb \gamma_{\rm SP}\Rb}{d \gamma_{\rm SP}}\Big{|}_{\gamma_{\rm SP}\,\,=\,\,\gamma_{cr}\Lb \zeta\Rb}\,\bas Y\,\,+\,\,\xi\,\,\,=\,\,\lambda\Lb \zeta\Rb \,\bas\,Y\,\,+\,\,\xi; ~~~\mbox{where}\,\,\xi\,\,=\,\,\ln\Lb r^2 \,Q^2_s\Lb Y=Y_0, b=0\Rb\Rb
\eeq
  we obtain    
      
    \beq \label{DSR2}   
    \frac{\partial}{\partial Y}\,=\,   \frac{\partial}{\partial\, z}  \frac{\partial  \,z }{\partial Y} \,=\,\bas \Big( \lambda\Lb \zeta\Rb\,\,-\,\,\zeta  \frac{d \lambda\Lb \zeta\Rb}{d \zeta}\Big) \frac{d}{d \,z}\,\,\equiv\,\,\bas\,\kappa\Lb \zeta\Rb\,\frac{d}{d \,z}    \eeq 
    
    Introducing
\beq \label{BKLT1}
\frac{ \partial \widetilde{N}\Lb Y; \xi, \vec{b}\Rb  }{\partial \xi} \,\,=\,\,1 - e^{- \phi\Lb Y,\xi,\vec{b}\Rb}
\eeq     
and searching for the solution that depends on $z$,   
  we can re-write \eq{BK2} in the form:
      \beq \label{BKLT5}
 \frac{d^2 \phi\Lb z,  \zeta\Rb}{d z^2} \,=\,\frac{1}{\kappa(\zeta)}\Lb 1\,\,-\,\,e^{ - \phi\Lb z,  \zeta\Rb}\Rb
 \eeq
     Introducing $\frac{d \phi\Lb \zeta\Rb}{d \zeta} \,\,=\,\, F\Lb \phi\Rb $
  we  can re-write \eq{BKLT5} in the form
 \beq \label{BKLT6}
  \h \frac{ d F^2\Lb \phi\Rb}{d \phi}\,=\,\frac{1}{\kappa\Lb \zeta\Rb}\Lb 1
 - e^{ - \phi}\Rb;~~~~~~~
 F^2\Lb \phi\Rb = \int d \phi \frac{2}{\kappa\Lb \zeta\Rb}\Lb 1 - e^{ - \phi}\Rb \,=\,    \frac{2}{\kappa\Lb \zeta\Rb} \Big(-1 +  \phi + e^{- \phi} \, +\, C\Lb Y,b\Rb\Big)
 \eeq
 Finally,
 \beq \label{BKLT7}
  \sqrt{\h \kappa\Lb\zeta\Rb} \int^\phi_{\phi_0} \frac{d \phi'}{\sqrt{ -1 + \phi' + e^{-\phi'} + C\Lb \zeta\Rb}}\,\,=\,\,z + {\rm Const}
 \eeq
 The equations of \eq{BKLT7}-type are discussed in Ref.\cite{MATH} 
(see 
formula {\bf 4.1.1.}).  $\phi_0$ denotes the value of $\phi $ at $\zeta = 
0$.
From \eq{BKLT7} we see that  ${\rm Const}$ is equal to 0. We need to
 find $\phi_z\Lb z, \zeta\Rb|_{z=0}$ for matching with the linear 
evolution,
 which is given by \eq{SP1} - \eq{SP2}. These equations for the kernel
 of \eq{SIMKER} can be re-written in the form:
   \begin{subequations}  
  \bea
 - \frac{\bas}{\gamma_{\rm cr}^2}\,Y\,+\,\,\xi\,\,&=&\,\,0\label{SEQ1}\\
  \frac{\bas}{\gamma_{\rm cr}}\,Y  \,\,-\,\,\Lb 1\,-
\,\gamma_{\rm cr}\Rb \xi   \,\,\,&=&\,\,4\,(m\,+\,m_0) \,b;\label{SEQ2}
     \eea 
 \end{subequations}
 which leads to the solution for $\gamma$ and the expression
 for the saturation momentum:
  \beq \label{DSR9}
  \gamma_{cr} \,\,=\,\,\frac{1}{2\,-\,\zeta};~~~~~~~~~~~~~
  \xi_s\,\,=\,\Lb 2\,-\,\zeta\Rb^2\,\bas\,Y\,\,\equiv
\,\,\lambda\Lb \zeta\Rb \,\bas\,Y;\,\,~~~~~~~~\kappa\Lb \zeta\Rb\,
\,=\,\,4 \,-\,\zeta^2
  \eeq
     
     Therefore,  for $\zeta \,>\,1$ , \eq{SEQ2} does not have a
 solution resulting in the scattering amplitude which  is  smaller 
than
 unity,  $N\Lb \xi', Y, b\Rb \,<\,1$. Hence, $\zeta =1$ gives the radius
 of the Froissart disc($R_F$) in  this case: $R_F\,=\,\bas\,Y/(4\,(m + m_0)$.
 
 One can see that for $\phi_0 \,<\,1$:
 \beq \label{BKLT8}
   \sqrt{ \kappa\Lb\zeta\Rb}\frac{1}{\sqrt{ \phi^2_0 \, +\,2\, C\Lb \zeta\Rb}}\,\frac{\partial\, \phi\Lb z, \zeta\Rb}{\partial \,z}|_{z=0 + \epsilon}\,\,=\,\,1
   \eeq
   For linear evolution at $z\,<\,0$ we have 
   \beq \label{BKLT9}
   \frac{\partial\, \phi\Lb z, \zeta\Rb}{\partial \,z}|_{z = 0 - \epsilon}  \,\,=\,\,\Lb 1\,-\,\gamma_{\rm cr}\Rb\,\phi_0;~~~~~~~~~ \phi\Lb z = 0 - \epsilon , \zeta\Rb\,\,=\,\,\phi_0;
 \eeq
 Hence the matching condition has the following form:
   \beq \label{BKLT91}
   \frac{\partial\, \phi\Lb z, \zeta\Rb}{\partial \,z}|_{z = 0 + \epsilon}  \,\,=\,\,   \frac{\partial\, \phi\Lb z, \zeta\Rb}{\partial \,z}|_{z = 0 - \epsilon}  \,\,= 
   \,\,\Lb 1\,-\,\gamma_{\rm cr}\Rb\,\phi_0;~~ \phi\Lb z = 0 + \epsilon , \zeta\Rb\,\,=\,\,\phi\Lb z = 0 - \epsilon , \zeta\Rb\,\,=\,\,\phi_0;~~\mbox{ at}~~~\epsilon\,\to\,0
 \eeq 
 
   Plugging this equation in \eq{BKLT8} we obtain
     \beq \label{BKLT10}
   C\Lb \zeta\Rb\,\,=\,\,\h \phi^2_0\,\Lb\kappa\Lb \zeta\Rb\,\Lb 1 \,-\,\gamma_{cr}\Rb^2\,-\,1\Rb\,\,=\,\,\h \phi^2_0\,   \Bigg(\Big(1 \,+\,\frac{\zeta}{\sqrt{1\,-\,\zeta}\,\Lb 1\,+\,\sqrt{1\,-\,\zeta}\Rb}\Big) \,\Lb 1\,-\,\zeta\Rb\,\,-\,\,1\Bigg)
   \eeq

   The explicit form of the solution at $\phi\,\to\,\phi_0$ takes the form:
      \bea \label{BKLT11}   
 &&   \phi\Lb z, \zeta\Rb\,\,=\\
 &&\,\,\h \phi_0\,\Bigg(   \Big( 1\,+\,\sqrt{\kappa\Lb \zeta\Rb}\,(1\,-\,\gamma_{cr}\Lb \zeta\Rb)\Big)\,\exp\Lb \frac{z}{\sqrt{\kappa\Lb \zeta\Rb}}\Rb\,\,+\,\, \Big( 1\,-\,\sqrt{\kappa\Lb \zeta\Rb}\,(1\,-\,\gamma_{cr}\Lb \zeta\Rb)\Big)\,\,\exp\Lb -\,\frac{z}{\sqrt{\kappa\Lb \zeta\Rb}}\Rb \Bigg)  \nn
 \eea

\begin{figure}[ht]
     \begin{center}
     \includegraphics[width=9.5cm]{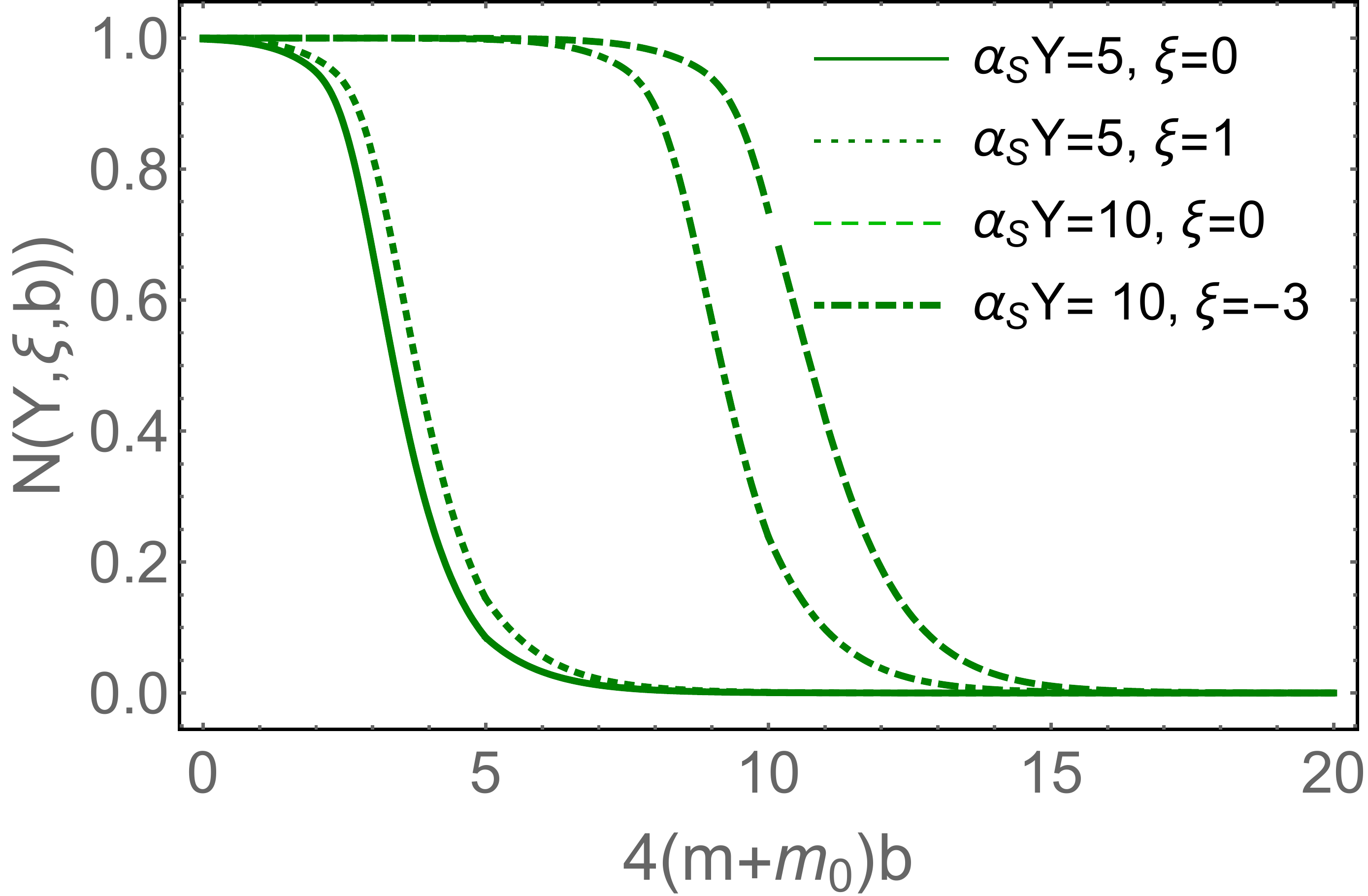}   
     \end{center}    
      \caption{ The solution to \eq{BKLT7} with the boundary 
conditions given by \eq{DSR9}. $\phi_0$ is taken  to be $0.01$.
      }
      \label{AM}
   \end{figure}

 For large $\phi$  the denominator in   \eq{BKLT7}  takes the
 form,  $\sqrt{\phi'}$ leading 
 to $\phi = {\rm Const} \exp\Lb -
 z^2/\Lb 2 \kappa\Lb\zeta\Rb\Rb\Rb $,   which is the scattering
 amplitude in the approach of Ref.\cite{LETU}
 for our simplified BFKL kernel.     In \fig{AM} we present
 the numerical solution to \eq{BKLT7}  which shows that 
 non-linear equation generates the impact parameter dependence 
which is typical for the Froissart  disc with radius 
($R_F$) proportional to $Y$ and  $|b - R| \,\propto 1/(m + m_0)$ .
  ~

           
        \section{Conclusions}                   
   
 In the paper we derived the generalization of the BFKL equation in 
 Gribov-Zwanziger  approach\cite{GRI0,GRI1,GRI2,GRI3,GRI4,GRREV,DOKH},  
to the confinement of quarks and gluons.  We found the solution to this
 modified BFKL equation at large impact parameters. This solution shows
 that generally speaking, this equation
 includes a
 dimensional scale, which provides the exponential decrease of the
 scattering amplitude at large impact parameters. Such behaviour of the 
scattering amplitude leads to the radius of interaction which  at 
high energies   increases 
as $\ln(1/x)=Y$ . Solving the non-linear  evolution 
equation  for deep inelastic scattering we  calculated the $x$ and $r$
 dependence of this radius.

  However , it turns out that  for  the Gribov propagator (see 
\eq{GLPR})
 of the gluon, which  tends to zero at small momenta ($G(q) \,
\,\xrightarrow{q \,\ll\,\mu} \,\, q^2/\mu^4$), the solution to
 the modified BFKL equation does not show an  exponential decrease
 , leading to the scattering amplitude that decreases as a power of
 the impact parameter.  Fortunately, for the general form of the gluon
 propagator in the  Gribov-Zwanziger  approach, in which the gluon 
propagator
 is finite at small momenta  ($G(q)\, \,\xrightarrow{q \,\ll\,\mu} \,\,
 (m^2 + m^2_0)/( m^4\,+\,\mu^4)$), we have indeed  an  exponential
 decrease. It should be emphasized, 
 that only such a  gluon propagator can be in accord with the lattice
 QCD estimates\cite{DOS}.

  We discuss the solution to a new equation, and single out the problem
 that the behaviour of the  intercept of the BFKL Pomeron, estimated in
 the variational approximation,  does not follow our expectations that 
we obtain on  general grounds  from \eq{HGR6}. Indeed, the general discussion in the
 spirit of Ref.\cite{LLS} leads to a small intercept in the case of
 the Gribov gluon propagator, and to the same intercept as for the 
massless
 BFKL Pomeron, in the case that describes the lattice QCD 
results\cite{DOS}.
 The variation approximation, developed in the paper, leads to the
 intercept of the massless BFKL Pomeron for the Gribov's  gluon propagator
 and  a sufficiently smaller intercept for the realistic case. We consider 
 as
 the next topic for us, is to find  the numerical solution for the 
spectrum 
of
 the suggested equation.

 In the paper we  have investigated the  impact parameter dependence of 
the 
solutions to the master equation in the
 entire kinematic region of impact parameters, without
 the  additional assumption that the  variable $\zeta\,\,\ll\,\,1$(see 
\eq{GACR}).
 
 We hope that this paper demonstrates, why and how the suggested
modified non-linear equation resolves the main difficulty
of the CGC approach: power-like decrease of the solution at large values of
 the impact parameter; and clarifies the physical meaning of the 
 non-perturbative dimensional scale, which was introduced
 in addition to the saturation scale.

  {\it Acknowledgements.} \\
   We thank our colleagues at Tel Aviv University and UTFSM for
 encouraging discussions. The special thanks go to Marat Siddikov for
 his advices on numerical estimates of the variational approximation.
 This research was supported  by 
 ANID PIA/APOYO AFB180002 (Chile)  and  Fondecyt (Chile) grant 1180118 .  
 
 \appendix
 \begin{boldmath}
\section{$\Sigma\Lb \kappa\Rb$}
\end{boldmath}

Using \eq{GGLPR3} we can re-write \eq{GGLTR2} in the form:
\beq \label{A1}
\Sigma\Lb \kappa\Rb\,\,=\,\,\h {\rm Re}\Bigg( \int\,d \kappa' \Big\{ \Lb 1\,\,+\,\,m_0\,i\Rb^2 \,G^+\Lb \vec{\kappa} - \vec{\kappa}'\Rb\,G^+\Lb \vec{\kappa}'\Rb\,\,+\,\,\Lb 1\,\,+\,\,m^2_0\Rb\,G^+\Lb \vec{\kappa} - \vec{\kappa}'\Rb\,G^-\Lb \vec{\kappa}'\Rb\Big\}\Bigg)
\eeq
Using \eq{SIGMA} and plugging in this equation $m_1\,= m_2\, =
\sqrt{m\,+\,i}\,=\,\sqrt{m\,+\,1}\,\,e^{ \h i \tan^{-1}\Lb 
\frac{1}{m}\Rb}$   we obtain :

\beq \label{A2}
 I_1\Lb m, \kappa\Rb\,\,=\,\,\int d \kappa' G^+\Lb \vec{\kappa} - \vec{\kappa}'\Rb\,G^+\Lb \vec{\kappa}'\Rb\,\,=\\
\displaystyle{ \frac{2}{\sqrt{\kappa  (\kappa +4 (m+i))}}  \ln \left(\frac{\sqrt{\kappa }+\sqrt{\kappa +4 (m+i)}}{-\sqrt{\kappa }+\sqrt{\kappa +4 (m+i)}}\right)}
 \eeq
 For $\int d \kappa' G^+\Lb \vec{\kappa} - \vec{\kappa}'\Rb\,G^-\Lb
 \vec{\kappa}'\Rb$ we have 
 \bea \label{A3}
&&I_2\Lb m, \kappa\Rb \,\,=\,\,\int d \kappa' G^+\Lb \vec{\kappa} - \vec{\kappa}'\Rb\,G^-\Lb \vec{\kappa}'\Rb\,\,=\\
&&\displaystyle{-\frac{1}{\sqrt{4 m \kappa+\kappa^2-4}}
 \ln\Lb \frac{ \kappa\,+\,2\,m\,-\,\sqrt{4 m \kappa+\kappa^2-4}}{\kappa\,+\,2\,m\,+\,\sqrt{4 m \kappa+\kappa^2-4}}\Rb}
 \eea
 
     \begin{figure}[ht]
     \begin{center}
     \begin{tabular}{c c c}
     \includegraphics[width=8.cm]{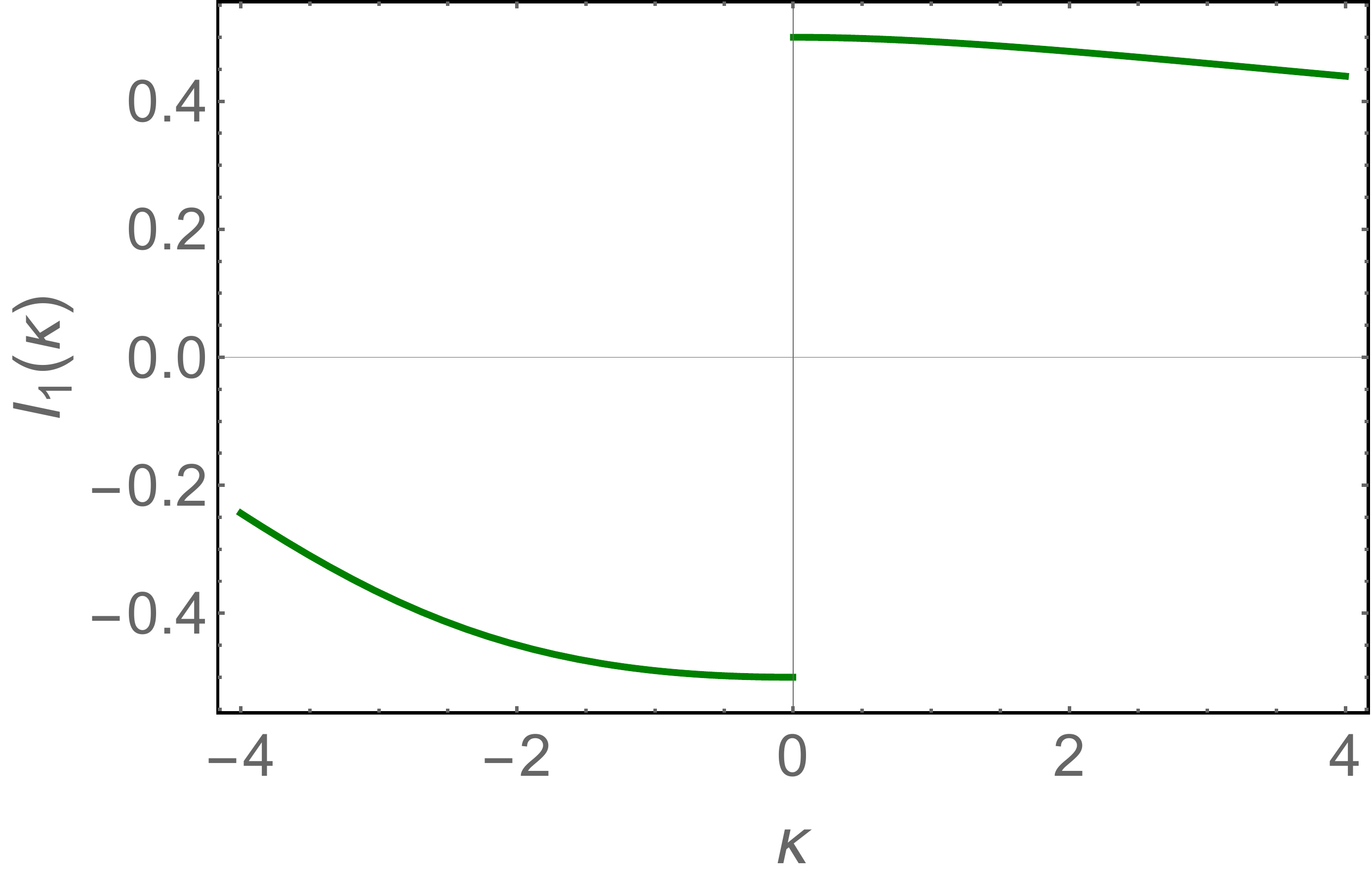} &~~~~&      \includegraphics[width=8.cm]{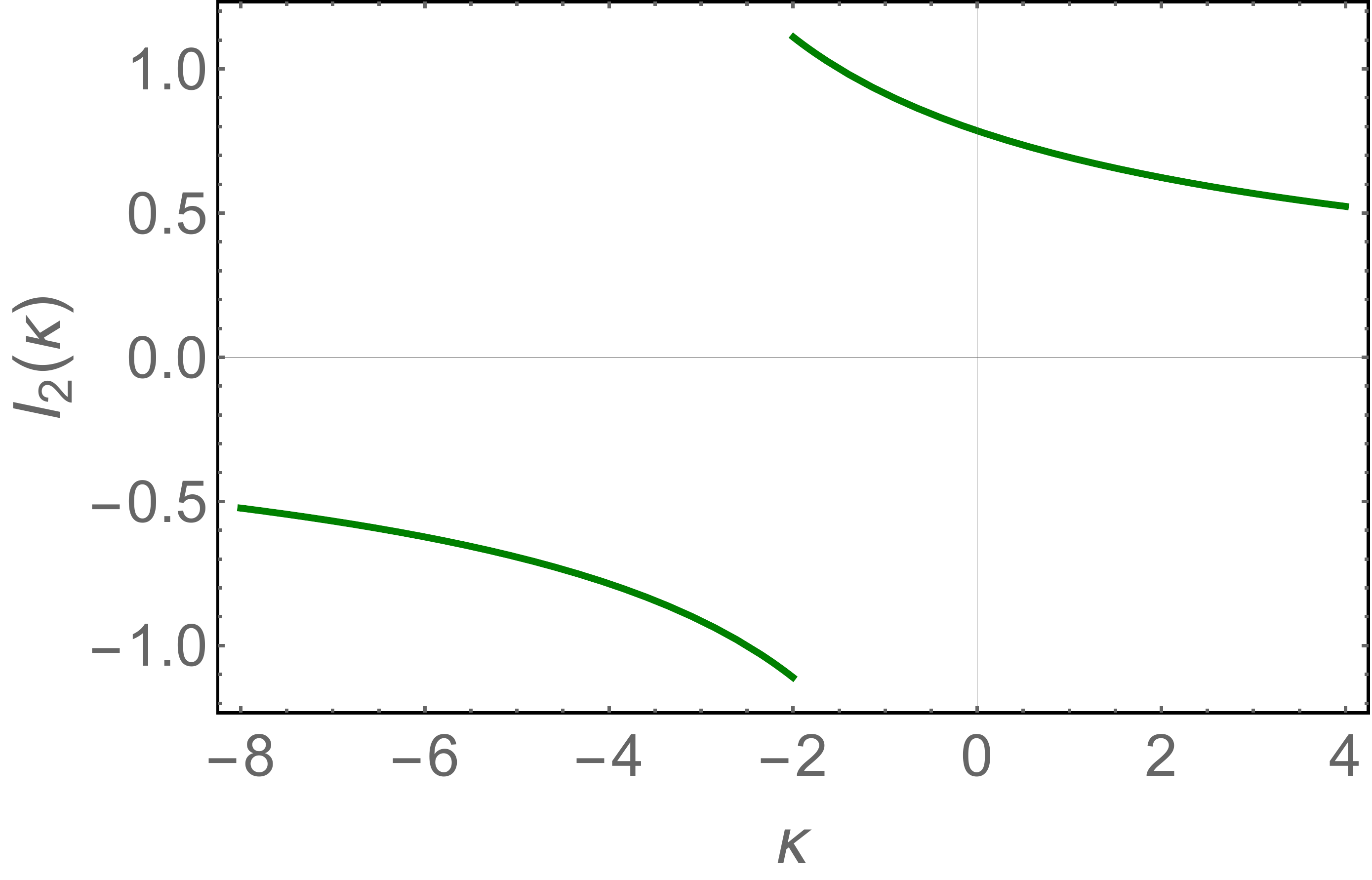}\\
     \fig{I}-a & &\fig{I}-b\\
     \end{tabular}
     \end{center}    
      \caption{$I_1\Lb m, \kappa\Rb$ ( \fig{I}-a) and $I_2\Lb m, 
\kappa\Rb$ ( \fig{I}-b) versus $\kappa$ at  $m^2$=1.}
\label{I}
   \end{figure}
  
In \fig{I} we plot the ${\rm Re}I_1$ and $I_2$ as function of 
 $\kappa$ at $m^2=1$. The singularities of $ I_1$ are  easy to see 
from
 the explicit expression in \eq{A2}:      $\kappa = 0$ and
 $\kappa = 4(m\,+\,i)$. However, the second singularities 
  for  the  $\kappa$ dependence  of ${\rm Re}I_1$  are not obvious. 
The
 possible singularities of $I_2$ stem from the solution of the equation:
\beq \label{A4}
4 m \kappa+ \kappa^2-4 =0;~~~ \kappa_{\pm}\,\,=\,\,2\Lb - m \pm \sqrt{m^2 + 1}\Rb
\eeq
However, it is easy to see that, both  $\kappa_+$ and $\kappa_-$ do
 not correspond to the singularity of $I_2$ (see \fig{I}-b).

Plugging \eq{A2} and \eq{A3} into \eq{A1} we obtain

\beq \label{A4}
\Sigma\Lb \kappa\Rb\,\,=\,\,\frac{1}{4} {\rm Re}\Bigg( \Big\{ \Lb 1\,\,+\,\,m_0\,i\Rb^2 \,I_1\,\,+\,\,\Lb 1\,\,+\,\,m^2_0\Rb  \,I_2\Bigg)
\eeq
 where $m$ and $m_0$ are defined as in  \eq{BFKLMR1}    - \eq{BFKLMR2}.
 \begin{boldmath}
\section{The numerical estimates in the variational approach.}
\end{boldmath}


As it has been pointed out in Ref.\cite{LLS}\footnote{We thank Marat
 Siddikov for the  instructions  for  obtaining  numerical estimates 
that he
 provided us. in private communications.} in  the integrals in \eq{VP2}
  we have two problems: (i) the very large values of $\kappa$ and
 $ \kappa'$ are essential; and (2) the region of $\kappa \,\to\,\kappa'$ 
 is very sensitive to the   numerical calculation  proceedure. 
Recall that
 this region in the case of massless BFKL equation leads to a divergency.
 To heal all these problems we re-write \eq{BFKLMR1} and \eq{VP2} in the 
following form:
\beq \label{B1}
E \,\phi\Lb \kappa\Rb\,\,=\,\,\, -\, T\Lb \kappa\Rb\,\phi\Lb \kappa\Rb\,-\,\,\bas \int d \kappa' \,K\Lb \kappa ,\kappa'\Rb\,\Bigg\{\phi\Lb \kappa'\Rb \,\,-\,\,\frac{G\Lb \kappa'\Rb}{G\Lb  \kappa\Rb} \,\phi\Lb \kappa\Rb\Bigg\}
+
\,\,m\,\frac{N^2+1}{N^2_c}G\Lb \kappa\Rb\int d \kappa' G\Lb \kappa'\Rb\,\phi\Lb \kappa\Rb
\eeq
where  $G\Lb \kappa\Rb$ is given by \eq{GGLPR1}. One can see that
 at $\kappa' \,\to\,\kappa$ the term in curly brackets vanishes,
 providing the smooth integration in this dangerous region. For a
better control of the interation at large $\kappa$ in \eq{VP2} we
 replace
\beq \label{B2}
T\Lb \kappa\Rb\,\,\to\,\,\Bigg( T\Lb \kappa\Rb\,\,-\,\,\ln \kappa\Bigg)\,\,+\,\,\ln \kappa
\eeq
The term in parentheses vanishes at large $\kappa$, leading to a
 converged integral at large $\kappa$, while the integral for the 
trial function of \eq{TRIF} can be taken, leading to the following expression:
\beq \label{B3}
\int d \kappa \Lb \ln \kappa\Rb\,\, \phi^2_{\rm trial}\Lb \kappa\Rb\,
\,=\,\,-\frac{\sqrt{\pi } 4^{-\gamma} \left(\frac{1}{a^2}\right)^{\frac{1}{2}
- \gamma} \left(a^2\right)^{1-2 \gamma} \Gamma \left(\gamma+\frac{1}{2}\right)
 \left((2 \gamma-1) \log \left(\frac{1}{a^2}\right)-2\right)}{(1-2 \gamma)^2
 \Gamma (\gamma)}
\eeq

All numerical integration were take replacing $\kappa = e^{l}$ and $\kappa'
 = e^{l'}$  and $l (l')$ runs from $-20$ to $250$.

 \end{document}